\documentclass[10pt]{article}
\usepackage{bbm}
\usepackage[final]{graphics}
\usepackage{amsmath}
\usepackage{amsfonts,amsbsy}
\usepackage{amssymb}

\def\empile#1\over#2{\mathrel{\mathop{\kern 0pt#1}\limits_{#2}}}

\newcommand{\slv}{\raise.15ex\hbox{$/$}\kern-.53em\hbox{$v$}}
\newcommand{\slF}{\raise.15ex\hbox{$/$}\kern-.53em\hbox{$F$}}
\newcommand{\slL}{\raise.15ex\hbox{$/$}\kern-.53em\hbox{$L$}}
\newcommand{\slP}{\raise.15ex\hbox{$/$}\kern-.53em\hbox{$P$}}
\newcommand{\slp}{\raise.15ex\hbox{$/$}\kern-.53em\hbox{$p$}}
\newcommand{\slq}{\raise.15ex\hbox{$/$}\kern-.53em\hbox{$q$}}
\newcommand{\slR}{\raise.15ex\hbox{$/$}\kern-.53em\hbox{$R$}}
\newcommand{\slQ}{\raise.15ex\hbox{$/$}\kern-.53em\hbox{$Q$}}
\newcommand{\slK}{\raise.15ex\hbox{$/$}\kern-.53em\hbox{$K$}}
\newcommand{\slk}{\raise.15ex\hbox{$/$}\kern-.53em\hbox{$k$}}
\newcommand{\slD}{\raise.15ex\hbox{$/$}\kern-.53em\hbox{$D$}}
\newcommand{\slC}{\raise.15ex\hbox{$/$}\kern-.53em\hbox{$C$}}
\newcommand{\slA}{\raise.15ex\hbox{$/$}\kern-.53em\hbox{$A$}}
\newcommand{\slSigma}{\raise.15ex\hbox{$/$}\kern-.53em\hbox{$\Sigma$}}
\newcommand{\slpartial}{\raise.15ex\hbox{$/$}\kern-.53em\hbox{$\partial$}}
\newcommand{\slcalP}{\raise.15ex\hbox{$/$}\kern-.63em\hbox{$\cal P$}}

\def\p{{\boldsymbol p}}
\def\q{{\boldsymbol q}}
\def\l{{\boldsymbol l}}
\def\k{{\boldsymbol k}}

\def\x{{\boldsymbol x}}
\def\y{{\boldsymbol y}}

\def\A{{\boldsymbol A}}
\def\B{{\boldsymbol B}}

\def\r{{\boldsymbol r}}

\catcode`\@=11

\newcount\@tempcntc
\def\@citex[#1]#2{\if@filesw\immediate\write\@auxout{\string\citation{#2}}\fi
  \@tempcnta\z@\@tempcntb\m@ne\def\@citea{}\@cite{%
        \@for\@citeb:=#2\do%
    {\@ifundefined{b@\@citeb}%
        {\@citeo\@tempcntb\m@ne\@citea%
                \def\@citea{,\penalty\@m\ }{\bf ?}\@warning%
                {Citation `\@citeb' on page \thepage \space undefined}}%
        {\setbox\z@\hbox{\global\@tempcntc0\csname b@\@citeb\endcsname\relax}%%
     \ifnum\@tempcntc=\z@ \@citeo\@tempcntb\m@ne%
       \@citea\def\@citea{,\penalty\@m}%
       \hbox{\csname b@\@citeb\endcsname}%
     \else%
      \advance\@tempcntb\@ne%
      \ifnum\@tempcntb=\@tempcntc%
      \else\advance\@tempcntb\m@ne\@citeo%
      \@tempcnta\@tempcntc\@tempcntb\@tempcntc\fi\fi}}\@citeo}{#1}}%

\def\@citeo{\ifnum\@tempcnta>\@tempcntb\else\@citea
  \def\@citea{,\penalty\@m}%
  \ifnum\@tempcnta=\@tempcntb\the\@tempcnta\else
   {\advance\@tempcnta\@ne\ifnum\@tempcnta=\@tempcntb \else
\def\@citea{--}\fi
    \advance\@tempcnta\m@ne\the\@tempcnta\@citea\the\@tempcntb}\fi\fi}

\catcode`\@=12

\begin{document}

%\begin{flushright}
%INT--PUB 06--?
%\end{flushright}

\title{\bf Drell-Yan production\\ and Lam-Tung relation in the\\ Color Glass Condensate formalism}
\author{Fran\c cois Gelis$^{(1)}$, Jamal Jalilian-Marian$^{(2,3)}$}
\maketitle
\begin{center}
\begin{enumerate}
\item Service de Physique Th\'eorique (URA 2306 du CNRS)\\
  CEA/DSM/Saclay, B\^at. 774\\
  91191, Gif-sur-Yvette Cedex, France
\item Institute for Nuclear Theory\\
  University of Washington\\
  Seattle, WA 98195-1550, USA
\item
  Department of Natural Sciences\\
  Baruch College\\
  New York, NY 10010, USA
\end{enumerate}
\end{center}

\begin{abstract}
We study the Drell-Yan production cross section and structure
functions in proton (deuteron)-nucleus collisions using the Color
Glass Condensate formalism.  The nucleus is treated in the Color Glass
Condensate framework which includes both higher twist effects due to
the inclusion of multiple scatterings and leading twist pQCD shadowing
due to the small $x$ resummation, while the proton (or deuteron) is
treated within the DGLAP improved parton model. In particular, the
Drell-Yan structure functions are used in order to investigate the
Lam-Tung relation at small $x$, which is known to be identically zero
at leading twist up to Next-to-Leading order, and is thus a good
playground for studying higher twist effects. In agreement with this,
we find that violations of this relation are more important for low
momentum and invariant mass of the Drell-Yan pair, and also in the
region of rapidity that corresponds to smaller values of $x$ in the
nucleus.

\end{abstract}
\vskip 5mm
\begin{flushright}
Preprint SPhT-T06/103
\end{flushright}

\section{Introduction}
The recent data from deuteron-gold collisions in the forward rapidity
region at RHIC \cite{Arsena1} have caused a great excitement in the
high energy heavy ion physics community. The observed suppression of
the hadron production transverse momentum spectrum in deuteron-gold
collisions as compared to proton-proton collisions (normalized to the
number of binary collisions) was qualitatively predicted by the
saturation physics and the Color Glass Condensate formalism
\cite{IancuLM1,IancuLM2,FerreILM1,IancuV1} and is now understood
quantitatively using saturation based models 
\cite{KharzKT1,KharzKT2,Jalil4,DumitHJ1,DumitHJ2}.

While the produced hadron spectra in deuteron-gold collisions in mid
rapidity can be described in terms of classical (Glauber like)
multiple scattering
\cite{KovchM3,KovneW1,KovchT1,DumitM1,DumitJ1,DumitJ2,GelisJ3,BlaizGV1,GelisM1,NikolS1,JalilK1},
one needs to go beyond a simple multiple scattering picture in order
to understand the suppression of the forward rapidity spectra. In the
Color Glass Condensate formalism, this suppression is due to the
evolution of the target nucleus wave function with $x$ (or
equivalently, with rapidity) described by the JIMWLK evolution
equations
\cite{JalilKLW1,JalilKLW2,JalilKLW3,JalilKLW4,KovneM1,KovneMW3,JalilKMW1,IancuLM1,IancuLM2,FerreILM1}
(or their large $N_c$ limit known as the Balitsky-Kovchegov (BK)
equation \cite{Balit1,Kovch3}).

To further probe the extend to which saturation is the dominant
physics in the forward rapidity region at RHIC, and in order to
clarify the role of the additional effects, such as parton
recombination, cold matter energy loss, etc., which have recently been
suggested \cite{HwaYF1,KopelNPJS1,QiuV1} as the reason for the
suppression pattern, it is essential to consider the electromagnetic
probes of the Color Glass Condensate, such as photon and dilepton
production  \cite{GelisJ1,GelisJ2,Jalil5} or photon-hadron 
correlations \cite{Jalil6}. The Color Glass Condensate
formalism would predict a similar suppression pattern for production
of photons and dileptons while the parton recombination models should
not since photons and dileptons do not interact strongly. Therefore,
observation of (or lack thereof) a similar suppression in the
production of photons (or dileptons) as in hadron production would
firmly establish saturation as the underlying physics of the phenomena
in the forward rapidity region at RHIC. In this paper we consider
dilepton production in proton-nucleus collisions and provide
quantitative predictions for Drell-Yan structure functions and angular
correlations in the kinematic region appropriate for the PHENIX
experiment at RHIC and CMS and ATLAS experiments at LHC \cite{Enter1}.

\section{Drell-Yan cross-section in the CGC formalism}
\label{sec:setup}
\subsection{Generalities}
Electromagnetic observables (such as photons and dileptons) are
cleaner probes of new phenomena than hadronic ones since they do not
interact strongly and are not sensitive to the non-perturbative, and
thus poorly understood, physics of hadronization. However they do
suffer from the fact that their production rates is much smaller than
the hadron production rate, due to smallness of the electromagnetic
coupling constant, and that in case of photons, measuring them
precisely poses new experimental difficulties.

When looking for new physics effects, it helps enormously when the
leading order processes which constitute the background vanish. This
is the case for the so called Lam-Tung relation \cite{LamT2,LamT3} in
dilepton production in perturbative QCD. It is known that the Leading
Order (LO) pQCD partonic cross sections satisfy the Lam-Tung
relation. This is the analog of the Callan-Gross relation between the
structure functions $F_1$ and $F_2$ in Deeply Inelastic
Scattering (DIS) which reads $F_2 \equiv 2\,x\, F_1$ or alternatively,
$F_L \equiv F_2 - 2\, x\, F_1 =0$. This is an indication that the
photon probe scatters off of a spin half object (a fermion) in the
target and as such, helped establish quarks as the constituents of the
target proton. The Callan-Gross relation is, however, not exact and
receives $\alpha_s$ corrections due to radiation of gluons such that
$F_L \sim \alpha_s \, xG (x,Q^2)$.

The Lam-Tung relation on the other hand does not receive any
$\alpha_s$ corrections (the first correction is of order $\alpha_s^2$)
and therefore can be a very sensitive probe of new (beyond standard
pQCD) physics, such as higher twists effects and/or a change from
DGLAP to BFKL kinematics signified by the change of the anomalous
dimension of the gluon distribution function. Here, we use the Color
Glass Condensate formalism, which includes both the multiple
scattering and BFKL anomalous dimension into effect, in order to
investigate the possible beyond the standard pQCD physics in the
Lam-Tung relation.

\subsection{The cross-section and hadronic tensor}
\begin{figure}[htb]
\begin{center}
\resizebox*{!}{2cm}{\includegraphics{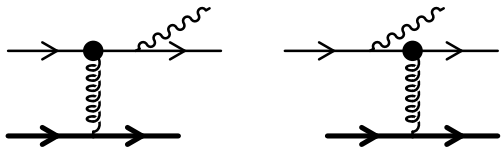}}
\end{center}
\caption{\label{fig:diagrams}The two diagrams contributing to the
emission of a photon in pA collisions. The black dot denotes the
Wilson line that resums all the multiple scatterings of the quark off
the color field of the nucleus.}
\end{figure}

In \cite{GelisJ1} and \cite{GelisJ2}, the expressions for photon and
dilepton production cross sections in proton (deuteron)-nucleus
collisions were derived, using a hybrid description in which the
proton is described by ordinary parton distributions while the nucleus
is described in terms of classical color sources and the color field
they produce.  In such a description, the leading process for the
emission of a photon (real or virtual) is a quark coming from the
proton, scattering off the classical color field of the nucleus, and
radiating a photon. Two terms contribute to the production amplitude,
depending on whether the photon is emitted before or after the
scattering of the quark off the nucleus, as illustrated in figure
\ref{fig:diagrams}.
There could potentially be a third diagram, where the emission of the
photon occurs inside the nucleus, so that the quark can interact with
the nucleus before and after the photon emission. However, such a
contribution is suppressed by the inverse of the collision energy due
to the Lorentz contraction of the nucleus.

If we denote $p$ the momentum of the incoming quark, $q$ the momentum
of the outgoing quark, and $k$ the momentum of the emitted photon, the
amplitude for photon emission in a given configuration of the target
color field reads \cite{GelisJ1,GelisJ2}~:
\begin{eqnarray}
&&
{\cal M}^\mu(\p|\q,k)
= 2\pi\delta(p^-\!-q^-\!-k^-)
\!\int\! d^2\x_\perp\; e^{i(\q_\perp+\k_\perp-\p_\perp)\cdot\x_\perp}\;
\Big[V(\x_\perp)-1\Big]
\nonumber\\
&&\qquad\qquad\qquad\qquad\times
ee_q\,\overline{u}(\p)\left\{
\frac{\gamma^-(\slq+\slk)\gamma^\mu}{(q+k)^2}
+
\frac{\gamma^\mu(\slp-\slk)\gamma^-}{(p-k)^2}
\right\}u(\q)\; .
\end{eqnarray}
In this formula, the nucleus is assumed to be moving in the positive
$z$ direction. We have neglected the mass of the quark, and $e_q$ is
the electrical charge of the quark in units of the electron charge
$e$. $V(\x_\perp)$ is a Wilson line defined as follows~:
\begin{equation}
V(\x_\perp)\equiv T_- \exp \left[-ig\int dz^-
A^+_a(z^-,\x_\perp)t^a\right]\; ,
\end{equation}
with $t^a$ the generators of the fundamental representation of SU(3)
and $A^+_a$ the classical color field of the nucleus in the color
glass condensate description. In terms of the number density of color
charges in the nucleus, $\rho_a(\x_\perp)$, its expression in the
covariant gauge is~:
\begin{equation}
A_a^+(x^-,\x_\perp)=-g\delta(x^-)\frac{1}{\partial_\perp^2} \rho_a(\x_\perp)\; .
\end{equation}

Turning this photon production amplitude into the amplitude for the
production of a lepton pair, where the lepton has momentum $k_1$ and
the anti-lepton has momentum $k_2$ ($k_1+k_2=k$), is
straightforward. One simply needs to multiply it by the free
propagator of a virtual photon, which in the Feynman gauge is
$D_{\mu\nu}(k)=-ig_{\mu\nu}/k^2$, and by the leptonic current
$J^\nu\equiv e\overline{v}(\k_2)\gamma^\nu u(\k_1)$. Therefore, the
Drell-Yan amplitude is given by~:
\begin{eqnarray}
&&
{\cal M}_{_{DY}}(\p|\q,\k_1,\k_2)
= 2\pi\delta(p^-\!\!-q^-\!\!-k^-)
\!\!\int\!\! d^2\x_\perp\, e^{i(q_\perp+\k_\perp-\p_\perp)\cdot\x_\perp}\,
\Big[V(\x_\perp)\!-\!1\Big]
\nonumber\\
&&\qquad\qquad\times
ie^2e_q\overline{u}(\p)\left\{
\frac{\gamma^-(\slq+\slk)\gamma^\mu}{(q+k)^2}
+
\frac{\gamma^\mu(\slp-\slk)\gamma^-}{(p-k)^2}
\right\}u(\q)\;\frac{\overline{v}(\k_2)\gamma_\mu u(\k_1)}{(k_1+k_2)^2}
\; .
\nonumber\\
&&
\end{eqnarray}

It will be convenient to introduce a notation for the part of this
amplitude that remains after we factor out the factor
$2\pi\delta(p^-\!\!-q^-\!\!-k^-)$~:
\begin{equation}
{\cal M}_{_{DY}}(\p|\q,\k_1,\k_2)
\equiv 2\pi\delta(p^-\!\!-q^-\!\!-k^-)
M_{_{DY}}(\p,\q;k_1,k_2)\; .
\end{equation}
Indeed, as was shown in \cite{GelisJ1}, the cross-section for the
process $qA\to q l^+l^-A$ is then given in terms of $M_{_{DY}}$ by the
following formula~:
\begin{eqnarray}
&&
d\sigma_{_{DY}}
=
\frac{1}{2p^-}
\frac{d^2\q_\perp dq^-}{(2\pi)^3 2q^-}\;
\frac{d^2\k_{1\perp}dk_1^-}{(2\pi)^3 2k_1^-}\;
\frac{d^2\k_{2\perp}dk_2^-}{(2\pi)^3 2k_2^-}\;
\nonumber\\
&&
\qquad\quad\times
2\pi\delta(p^--q^--k_1^--k_2^-)\;
\frac{1}{N_c}\sum_{\rm colors}
\left<\left|M_{_{DY}}(\p|\q,\k_1,\k_2)\right|^2\right>_{_A}
\; ,
\end{eqnarray}
where $E_q$, $E_1$, and $E_2$ are respectively the energy of the final
quark, of the lepton and of the anti-lepton. The bracket
$\big<\cdots\big>_{_A}$ denotes the average over the color sources in
the target nucleus. We have summed over the colors of the final quark
and averaged over the colors of the incoming quark.  When we square
the amplitude in order to obtain the cross-section, it is customary to
write it in the following way~:
\begin{equation}
d\sigma_{_{DY}}
=\frac{1}{(2\pi)^4}
\frac{\alpha_{\rm em}^2}{M^2} \;
\frac{d^2\k_{1\perp}dk_1^-}{k_1^-}\;
\frac{d^2\k_{2\perp}dk_2^-}{k_2^-}\;
e_q^2 W_{\mu\nu} L^{\mu\nu}\; ,
\label{eq:sigma-qA}
\end{equation}
where $M$ is the invariant mass of the lepton pair ($M^2\equiv
(k_1+k_2)^2$), and $L^{\mu\nu}$ is a leptonic tensor coming from the
trace over the Dirac indices in the lepton loop\footnote{We also
neglect the lepton mass.},
\begin{equation}
L^{\mu\nu}\equiv 
g^{\mu\nu}-\frac{k_1^\mu k_2^\nu+k_1^\nu k_2^\mu}{k_1\cdot k_2}\; ,
\end{equation}
and where $W_{\mu\nu}$ contains the factors that depend on the nuclear
target~:
\begin{eqnarray}
&&W_{\mu\nu}=\frac{1}{2p^-}\int\frac{d^2\q_\perp dq^-}{(2\pi)^2 q^-}
\; \delta(p^--q^--k^-)
\int d^2\x_\perp d^2\y_\perp
\; e^{i(\q_\perp+\k_\perp)\cdot(\x_\perp-\y_\perp)}
\nonumber\\
&&\;\times
\frac{1}{N_c}
{\rm tr}\,
\left<\Big[V(\x_\perp)\!-\!1\Big]\Big[V^\dagger(\y_\perp)\!-\!1\Big]\right>_{_A}
\!
{\rm tr}\,
\left\{
\slp\Big[
\frac{\gamma^-(\slq+\slk)\gamma_\nu}{(q+k)^2}
+
\frac{\gamma_\nu(\slp-\slk)\gamma^-}{(p-k)^2}
\Big]
\right.
\nonumber\\
&&\qquad\qquad\qquad\qquad\qquad\qquad\quad\times\left.
\slq\Big[
\frac{\gamma^-(\slp-\slk)\gamma_\mu}{(p-k)^2}
+
\frac{\gamma_\mu(\slq+\slk)\gamma^-}{(q+k)^2}
\Big]
\right\}
\; .
\label{eq:Wmunu}
\end{eqnarray}
In this formula, we denote $k\equiv k_1+k_2$, and we have assumed that
the incoming quark has no transverse momentum: $\p_\perp=0$. Of
course, the integration over the variable $q^-$ is trivially done
thanks to the delta function. Note that the first trace is over color
indices while the second one is over the indices carried by the Dirac
matrices.

By assuming translation invariance in the transverse plane for the
nucleus, this formula may be rewritten in terms of the Fourier
transform of the correlator of two Wilson lines~:
\begin{eqnarray}
  &&W_{\mu\nu}=\frac{\pi R_{_A}^2}{2p^-(p^--k^-)}
  \int\frac{d^2\q_\perp}{(2\pi)^2}\,C(\q_\perp+\k_\perp)
  \nonumber\\
  &&\;\qquad\qquad\times
  {\rm tr}\,
  \left\{
    \slp\Big[
    \frac{\gamma^-(\slq+\slk)\gamma_\nu}{(q+k)^2}
    +
    \frac{\gamma_\nu(\slp-\slk)\gamma^-}{(p-k)^2}
    \Big]
  \right.
  \nonumber\\
  &&\qquad\qquad\qquad\qquad\quad\times\left.
    \slq\Big[
    \frac{\gamma^-(\slp-\slk)\gamma_\mu}{(p-k)^2}
    +
    \frac{\gamma_\mu(\slq+\slk)\gamma^-}{(q+k)^2}
    \Big]
  \right\}
  \; .
\label{eq:Wmunu1}
\end{eqnarray}
where we denote
\begin{equation}
  \pi R^2_{_A} 
  C(\l_\perp)\equiv \int d^2\x_\perp d^2\y_\perp
  \;e^{i\l_\perp\cdot (\x_\perp-\y_\perp)}\;
  \frac{1}{N_c}{\rm tr}\,
  \left<V(\x_\perp)V^\dagger(\y_\perp)\right>_{_A}\; ,
\end{equation}
$R_{_A}$ being the radius of the nucleus. The function $C(\l_\perp)$
is directly related to the Fourier transform of the cross-section for 
a $q\bar{q}$ dipole scattering on a target nucleus. Therefore, it
could in principle be obtained from fits of Deep Inelastic Scattering
on the same target. Note also that we have dropped terms with no
Wilson lines, as well as terms with only one Wilson line. These terms
have a $\delta(\q_\perp+\k_\perp)$ under the integral, and correspond
to a situation where no transverse momentum is exchanged between the
target nucleus and the quark line. They do not contribute to the
photon production cross-section.

So far, we have written the Drell-Yan cross-section and structure
functions for an incoming quark colliding on a nucleus. Naturally,
they will be measured for incoming nucleons. The relation between the
quantities defined in eqs.~(\ref{eq:sigma-qA}) and (\ref{eq:Wmunu}) for
the quark-nucleus system and the equivalent quantities for the
proton-nucleus system are~:
\begin{eqnarray}
&&
d\sigma_{_{DY}}^{\rm pA}=\int_0^1dx_1 \; q(x_1,\mu^2)\; 
\left.d\sigma_{_{DY}}\right|_{p^-=x_1\sqrt{s/2}}\; ,
\nonumber\\
&&
W_{\mu\nu}^{\rm pA}=\int_0^1dx_1 \; q(x_1,\mu^2)\; 
\left.e_q^2\,W_{\mu\nu}\right|_{p^-=x_1\sqrt{s/2}}\; ,
\end{eqnarray}
where $x_1$ is the longitudinal momentum fraction of the quark in the
incoming nucleon. Note that the only combination of parton
distributions of the proton that matter in this calculation is the
structure function $F_2$~:
\begin{equation}
F_2(x_1,\mu^2)\equiv\sum_q e_q^2\, q(x_1,\mu^2)\; ,
\end{equation}
since the ``probe'' with which we explore the proton is a photon. Most
of our discussion will be about the structure functions of the
quark-nucleus subsystem, and we will convolute them with the quark
distribution of the proton (or deuteron in the case of RHIC) only at
the end when we present numerical results.

\subsection{Structure functions and the Lam-Tung relation}
\begin{figure}[htbp]
\begin{center}
\resizebox*{!}{5cm}{\includegraphics{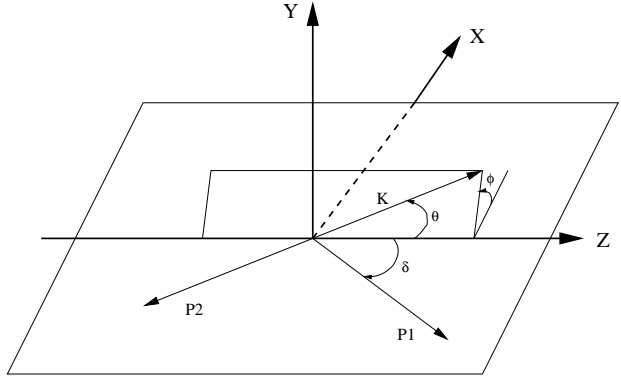}}
\end{center}
\caption{\label{fig:angles}Definition of the angle $\delta$.}
\end{figure}

It is customary to perform a decomposition of the hadronic tensor
$W_{\mu\nu}$ into four structure functions. There are a priori three
4-momenta upon which $W_{\mu\nu}$ might depend: $P_1^\mu, P_2^\mu$ the
momenta of the proton and of a nucleon in the nucleus respectively,
and $k^\mu$ the momentum of the virtual photon. This tensorial
decomposition is usually performed in the rest frame of the lepton
pair, where $\k=0$. Denoting $E_1$ and $E_2$ the energies of the
proton and of a nucleon in the nucleus {\sl in the rest frame of the
lepton pair}\footnote{The vectors $P_{1,2}^\mu$ should be kept in the
center of momentum frame of the proton-nucleon subsystem, since it is
in this frame that eq.~(\ref{eq:Wmunu}) for $W_{\mu\nu}$ is defined.},
one first defines the following four projectors\footnote{Note that,
since these projectors are defined in terms of the ratios
$P_1^\mu/E_1$ and $P_2^\mu/E_2$, they are identical for the
proton-nucleus system and for the quark-nucleus subsystem. This will
become more obvious with the explicit formulas in
eqs.~(\ref{eq:projs2}), where one can see that the center of mass
collision energy has disappeared. This means that we can perform these
projections on the $W_{\mu\nu}$ of the qA subsystem, and convolute
with the quark distribution of the proton only later.}~:
\begin{align}
&
P^{\mu\nu}_{_{TL}}\equiv -g^{\mu\nu}\; ,\quad
&&
P^{\mu\nu}_{_{L12}}\equiv \frac{P_1^\mu P_2^\nu+P_1^\nu P_2^\mu}{E_1E_2}\; ,
\nonumber\\
&
P^{\mu\nu}_{_{L1}}\equiv \frac{P_1^\mu P_1^\nu}{E_1^2}\; ,\quad
&&
P^{\mu\nu}_{_{L2}}\equiv \frac{P_2^\mu P_2^\nu}{E_2^2}\; .
\label{eq:projs1}
\end{align}
>From these projectors, one defines the following four structure functions~:
\begin{align}
&
T_{_{TL}}\equiv P^{\mu\nu}_{_{TL}} \; W_{\mu\nu}\; ,\quad
&&
T_{_{L12}}\equiv P^{\mu\nu}_{_{L12}} \; W_{\mu\nu}\; ,
\nonumber\\
&
T_{_{L1}}\equiv P^{\mu\nu}_{_{L1}} \; W_{\mu\nu}\; ,\quad
&&
T_{_{L2}}\equiv P^{\mu\nu}_{_{L2}} \; W_{\mu\nu}\; .
\end{align}
Alternatively, different sets of structure functions can be defined by
applying a $4\times 4$ ``rotation matrix'' to the previous set of
structure functions. The helicity structure functions can be obtained
in this way \cite{LamT1}. Several definitions of the helicity
structure functions exist in the literature, and in this paper we are
going to consider the so-called Collins-Soper structure functions,
defined as~:
\begin{equation}
\left[
\begin{matrix}
W_{_{TL}}\cr
W_{_L}\cr
W_{_\Delta}\cr
W_{_{\Delta\Delta}}\cr
\end{matrix}
\right]
\equiv
R_{_{CS}}\;
\left[
\begin{matrix}
T_{_{TL}}\cr
T_{_{L1}}\cr
T_{_{L2}}\cr
T_{_{L12}}\cr
\end{matrix}
\right]\; ,
\end{equation}
where the matrix $R_{_{CS}}$ is defined as~:
\begin{equation}
R_{_{CS}}\equiv 
\frac{1}{2}\;
\left[
\begin{matrix}
1 & 0 & 0 & 0
\cr\cr
0 & \frac{1}{2\cos^2\delta} & \frac{1}{2\cos^2\delta} & -\frac{1}{2\cos^2\delta}
\cr\cr
0 & -\frac{1}{\sin 2\delta} & \frac{1}{\sin 2\delta} & 0
\cr\cr
1 & \frac{1+\cos^2\delta}{\sin^2 2\delta}
& -\frac{1+\cos^2\delta}{\sin^2 2\delta}
& \frac{1-3\cos^2\delta}{\sin^2 2\delta}
\cr
\end{matrix}
\right]\; .
\label{eq:Rcs}
\end{equation}
In this formula, we use the so-called ``Collins-Soper frame'' in order
to define the angle $\delta$: in the lepton pair rest frame, the
momenta of the two incoming hadrons are not opposite, and the $Z$-axis
is chosen to bisect the angle between $P_1$ and $-P_2$. $\delta$ is the
angle between the $Z$-axis and the momentum $P_1$, defined in figure
(\ref{fig:angles}) and given by 
\begin{equation}
\cos^2\delta =\frac{M^2}{M^2+\k_\perp^2}\; ,\quad
\sin^2\delta =\frac{\k_\perp^2}{M^2+\k_\perp^2}\; .
\end{equation}

For the sake of completeness, we also show the expressions for the
structure functions explicitly,
\begin{eqnarray}
W_{_{TL}} &=& \frac{1}{2} T_{_{TL}}\; , \nonumber \\
W_{_L} &=& \frac{M^2+\k_\perp^2}{4M^2}\, 
\left[ T_{_{L1}} + T_{_{L2}} - T_{_{L12}}\right] \nonumber \\ 
W_{_{\Delta}} &=& \frac{1}{4} 
\frac{M^2+ \k_\perp^2}{M k_\perp} 
\left[T_{_{L2}} - T_{_{L1}}\right]\; , \nonumber \\
W_{_{\Delta\Delta}} &=& \frac{1}{2} T_{_{TL}} 
- \frac{ (M^2 +  \k_\perp^2) (2 M^2 +  \k_\perp^2)}
{8 M^2  \k_\perp^2}
\,\left[ T_{_{L1}} + T_{_{L2}}\right] \nonumber \\
&&\qquad\qquad+ 
\frac{ (M^2 +  \k_\perp^2) (- 2 M^2 +  \k_\perp^2)}{8 M^2  \k_\perp^2}
\,T_{_{L12}}\; .
\label{eq:Ws}
\end{eqnarray}
At leading twist, the following relation
\begin{eqnarray}
W_{_{\Delta\Delta}} - \frac{1}{2} W_{_L} &=&  
\frac{1}{2} \Bigg[ 
T_{_{TL}} 
- \frac{(M^2 + k_t^2)^2}{2 M^2 k_t^2} ( T_{_{L1}} + T_{_{L2}}) 
- \frac{M^4 -k_t^4}{2 M^2 k_t^2} T_{_{L12}}
\Bigg] \nonumber \\
&=& 0\; ,
\label{eq:lam-tung}
\end{eqnarray}
known as the {\sl Lam-Tung relation} \cite{LamT2,LamT3}, is valid up
to Next-to-Leading order, i.e. up to the order ${\cal O}(\alpha_s)$.

\subsection{Kinematics and matrix elements}
Below we will discuss violations of this relation in the kinematics
appropriate to the RHIC and LHC experiments. Before that, it is
instructive to be more specific.  The vectors $P_1^\mu$ and $P_2^\mu$
are proportional to unit vectors along the light cones\footnote{We
can afford to be a bit sloppy here about the use of the squared center
of mass energy for the proton-nucleus system, $s$, instead of that for
the quark-nucleus subsystem, $\hat{s}=x_1 s$. Indeed, since the
energies $E_1$ and $E_2$ are also proportional to $\sqrt{s}$, this
choice is irrelevant in the definition of the projectors.}~:
\begin{equation}
P_1^\mu=\sqrt{\frac{s}{2}}\;v_-^\mu\; ,\quad
P_2^\mu=\sqrt{\frac{s}{2}}\;v_+^\mu\; ,
\end{equation}
with
\begin{equation}
v_+^\mu\equiv\frac{1}{\sqrt{2}}(1,0,0,1)\; ,\quad
v_-^\mu\equiv\frac{1}{\sqrt{2}}(1,0,0,-1)\; .
\end{equation}
(Remember that the proton is moving in the negative $z$ direction and
the nucleus in the positive $z$ direction). The energies $E_1$ and
$E_2$ of the proton and of a nucleon of the nucleus are, in the rest
frame of the lepton pair,
\begin{equation}
E_1=\frac{e^{-y}}{2M}\;\sqrt{s(M^2+\k_\perp^2)}\; ,\quad
E_2=\frac{e^{+y}}{2M}\;\sqrt{s(M^2+\k_\perp^2)}\; ,
\end{equation}
so that the projectors of eqs.~(\ref{eq:projs1}) can be rewritten as~:
\begin{align}
&
P^{\mu\nu}_{_{TL}}\equiv -g^{\mu\nu}\; ,\quad
&&
P^{\mu\nu}_{_{L12}}\equiv \frac{2M^2}{M^2+\k_\perp^2}\,
\left[v_+^\mu v_-^\nu+v_+^\nu v_-^\mu\right]\; ,
\nonumber\\
&
P^{\mu\nu}_{_{L1}}\equiv \frac{2M^2e^{+2y}}{M^2+\k_\perp^2}\,v_-^\mu v_-^\nu\; ,\quad
&&
P^{\mu\nu}_{_{L2}}\equiv \frac{2M^2e^{-2y}}{M^2+\k_\perp^2}\,v_+^\mu v_+^\nu\; .
\label{eq:projs2}
\end{align}

Performing explicitly the Dirac algebra and the contractions with
these projectors, we obtain the following expression for the four
$T$'s~:
\begin{equation}
T_\alpha=\frac{\pi R_{_A}^2}{2(1-z)}
  \int\frac{d^2\q_\perp}{(2\pi)^2}\,C(\q_\perp+\k_\perp)\;{\cal T}_\alpha\; ,
\label{eq:T}
\end{equation}
where $\alpha \in \{TL,L1,L2,L12\}$ with
\begin{eqnarray}
&&
{\cal T}_{_{TL}} 
= 
16 \,\Bigg\{
\frac{(1-z) k_\perp^2}{[M^2 -2 p\cdot k]^2} 
+ 
\frac{[z \q_\perp - (1-z) \k_\perp]^2}{(1-z) [M^2 + 2 q\cdot k]^2} 
\nonumber\\
&&\qquad\qquad\qquad\qquad\qquad\qquad
-
2 
\frac{[(1-z)q_\perp^2+(2-z)\q_\perp\cdot\k_\perp]}
{[M^2-2p\cdot k][M^2+2q\cdot k]}\; 
\Bigg\} , 
\nonumber \\
&&
\nonumber \\
&&
{\cal T}_{_{L1}}
= 
\frac{32(1-z)}{z^2}\, 
{ M^2 } 
\bigg\{
\frac{1}{M^2 + 2 q\cdot k}
+ 
\frac{ 1-z}{M^2 -2 p\cdot k}\bigg\}^2\; ,
\nonumber \\
&&
\nonumber \\
&&
{\cal T}_{_{L2}}
=
\frac{32 z^2}{1-z}\,
\frac{ M^2}{ (M^2 + \k_\perp^2)^2} 
\bigg\{ 
\frac{q_\perp^2 (\q_\perp + \k_\perp)^2}{[M^2 + 2 q\cdot k]^2} \bigg\}\; , 
\nonumber \\
&&
\nonumber \\
&&
{\cal T}_{_{L12}}
=
64 \,
\frac{ M^2}{ M^2 + \k_\perp^2} 
\frac{\q_\perp \cdot (\q_\perp + \k_\perp)}{ [M^2 + 2 q\cdot k]} 
\bigg\{ 
\frac{1-z}{ M^2 - 2 p\cdot k} + 
\frac{1}{M^2 + 2 q\cdot k} 
\bigg\}\; .
\nonumber\\
&&
\label{eq:T-all}
\end{eqnarray}
Eqs.~(\ref{eq:T}) and (\ref{eq:T-all}) thus provide the full leading
result for the production of a lepton pair in proton-nucleus
collisions in the CGC framework. These results are for a quark with a
fixed longitudinal momentum $p^-$. In order to obtain the
corresponding results for an incoming proton, one needs simply to
convolute with the quark distribution $q(x_1)$, and use
eqs.~(\ref{eq:T-all}) with $z=k^-/(x_1\sqrt{{s}/{2}})$.

This result is of the same order in $\alpha_s$ as the standard NLO
$qg\to q\gamma^*\to q\ell^+\ell^-$ contribution, but compared to the
standard pQCD result it resums all the multiple scatterings on the
dense nucleus. One point should be emphasized: our calculation
incorporates also the ``LO'' process $q\bar{q}\to \gamma^*\to
\ell^+\ell^-$, although in a non obvious way. Indeed, the intermediate
-- virtual and space-like -- quark in the diagrams of figure
(\ref{fig:diagrams}) must be interpreted in some kinematical domain as
an antiquark belonging to the wave-function of the nucleus (in that
case, the quark in the final state came also from the nucleus
wave-function). Note however the pQCD NLO process $q\bar{q}\to
g\gamma^*\to g\ell^+\ell^-$ would only appear at the next order in our
calculation, via diagrams similar to those of figure
(\ref{fig:diagrams}) but with a gluon in the final state in addition to
the photon. In order to understand why this process does not appear on
the same footing here, one has to remember that in our approach the
antiquarks are produced explicitly from the gluons, and that a process
that has both an antiquark in the initial state and a gluon in the
final state is non-leading in our framework.

\subsection{The collinear factorization limit}
In eq.~(\ref{eq:Wmunu1}), the function $C(\q_\perp+\k_\perp)$ resums
all the multiple scatterings that the quark undergoes while it goes
through the nucleus. However, it is possible to extract from this
formula the contribution from a single scattering, i.e. the
leading-twist contribution. In this limit, one should recover the
results from collinear factorization in perturbative QCD.

This is done as follows. One assumes that the momentum scales
$\q_\perp,\k_\perp,M$ that characterize the final state are much
larger than the saturation scale, and one makes the approximation that
the transverse momentum of the incoming gluons,
$\l_\perp=\q_\perp+\k_\perp$, is large compared to $Q_s$ but small
compared to the scales of the final state. This approximation is
legitimate in this regime, because when there is a large range between
$Q_s$ and the scales of the final state, the integration over
$\l_\perp$ is dominated by a large logarithmic contribution that comes
from values of $\l_\perp$ comprised in this range. One must first
approximate the matrix elements ${\cal T}_\alpha$ by assuming that the
transverse momentum of the gluons coming from the nucleus,
$\l_\perp\equiv\q_\perp+\k_\perp$, is smaller than the other scales.
As pointed out after eq.~(\ref{eq:Wmunu1}), these matrix elements
vanish when $\l_\perp=0$, and therefore one should keep higher orders
in the expansion in $\l_\perp$. Formally, one can write~:
\begin{equation}
{\cal T}_\alpha=\A^i_\alpha\; \l_\perp^i+\B_\alpha^{ij}\; \l_\perp^i
\l_\perp^j+\cdots
\label{eq:Texp1}
\end{equation}
Since the function $C(\l_\perp)$ depends only on the modulus of the
vector $\l_\perp$ but not on its orientation, the term linear in
$\l_\perp$ in this expansion will vanish when inserted into
eq.~(\ref{eq:T}) when we perform the integration over the orientation
of the vector $\l_\perp$. We therefore need to expand the ${\cal
T}_\alpha$'s to second order in $\l_\perp$. Assuming this expansion
has been performed, we can now write
\begin{equation}
T_\alpha\approx
\frac{\pi R_{_A}^2}{4(1-z)}
B^{ii}_\alpha
\int\frac{d^2\l_\perp}{(2\pi)^2}\,l_\perp^2C(l_\perp)\; ,
\label{eq:Texp}
\end{equation}
Recalling the fact that in the leading-twist limit -- valid here since
$l_\perp \gg Q_s$ --, $l_\perp^2 C(l_\perp)$ is proportional to the
non-integrated gluon distribution of the nucleus, the integral over
$\l_\perp$ that appears in the previous equation gives the usual
integrated gluon distribution, $xG_{_A}(x,Q^2)$, with a resolution
scale $Q^2$ which is the upper limit of the range of validity of the
expansion in eq.~(\ref{eq:Texp1}).

It is straightforward to perform explicitly the expansion in powers of
$\l_\perp$ of the matrix elements ${\cal T}_\alpha$, in order to
obtain the second order coefficients $B^{ii}_\alpha$. One gets~:
\begin{eqnarray}
&&
B^{ii}_{_{L1}} 
=
128\, z^2 \, (1-z)^3 
\frac{M^2\, \k_\perp^2}{[\k_\perp^2 + (1-z) M^2]^4}\; , 
\nonumber\\
&&
\nonumber\\
&&
B^{ii}_{_{L2}}
=
 64\, z^4 \, (1-z)\, 
\frac{M^2\, \k_\perp^2}{(\k_\perp^2 + M^2)^2 \, [\k_\perp^2 + (1-z) M^2]^2}\; ,
\nonumber \\
&&
\nonumber\\
&&
B^{ii}_{_{L12}}
=
 -128 \, z^3\, (1-z)^2\, 
\frac{M^2 \, \k_\perp^2}{(\k_\perp^2 + M^2) \,  [\k_\perp^2 + (1-z) M^2]^3}\; ,
\nonumber \\
&&
\nonumber\\
&&
B^{ii}_{_{TL}}
=
 32 \, z^2 \, (1-z)\, \frac{1}{[\k_\perp^2 + (1-z) M^2]^4}
\nonumber\\
&&
\qquad\qquad
\times
\left[ 4 (1-z)^2 \k_\perp^2 M^2 
+ [1 + (1-z)^2][\k_\perp^4 + (1-z)^2 M^4]\right]
\; .
\end{eqnarray}
Using these expressions, it is easy to verify that the Lam-Tung sum
rule given in eq.~(\ref{eq:lam-tung}) holds identically in the leading
twist (single scattering) limit.

\subsection{Angular coefficients}
\label{sec:angular}
Based on the general Lorentz structure of the cross section, one can
write the transverse momentum integrated Drell-Yan cross section in
terms of the polar and azimuthal angles of the dilepton pair in the
dilepton center of mass frame as
\begin{equation}
\frac{dN}{d\Omega} \equiv \left[\frac{d\sigma}{d^4 k}\right]^{-1} \, 
\frac{d\sigma}{d\Omega d^4 k}
\label{eq:dndomega}
\end{equation}
where $d\Omega \equiv d\cos \theta \, d \phi$ is the solid angle, with
$\theta$ and $\phi$ defined in Fig. (\ref{fig:angles}) and $k$ is the 
dilepton four momentum with
\begin{eqnarray}
\frac{dN}{d\Omega} &=& \frac{3}{16 \pi W_{_{TL}}} 
\Big[
W_{_{TL}} (1 + \cos^2 \theta ) 
+ \frac{1}{2} W_{_L} (1 - 3 \cos^2 \theta ) 
\nonumber\\
&&\qquad\qquad\quad
+ W_{_\Delta} \sin 2 \theta \cos \phi 
+ W_{_{\Delta \Delta}} \sin^2 \theta \cos 2 \phi
\Big]\; .
\label{eq:def-ang}
\end{eqnarray}
It is customary to parameterize this cross-section in terms of the angular
coefficients $A_0, A_1, A_2$ as follows,
\begin{eqnarray}
\frac{dN}{d\Omega} &=& \frac{3}{16 \pi} 
\Big[1 + \cos^2 \theta + 
\frac{1}{2} A_0 \, (1- 3 \cos^2 \theta ) 
\nonumber\\
&&\qquad\qquad\quad
+ A_1 \, \sin 2 \theta \cos \phi 
+ \frac{1}{2} A_2 \, \sin^2 \theta \cos 2 \phi 
\Big]\; ,
\end{eqnarray}
so that 
\begin{equation}
A_0 \equiv \frac{W_{_L}}{W_{_{TL}}}\; , 
\;\;\;
A_1 \equiv \frac{W_{_\Delta}}{W_{_{TL}}} \; , 
\;\;\;
A_2 \equiv \frac{2 W_{_{\Delta\Delta}}}{W_{_{TL}}} 
\label{eq:ang}
\end{equation}
in terms of which the Lam-Tung relation simply reads $A_0 =
A_2$. Below we will show our results for the structure functions $W$'s
as well as the angular parameters $A$'s.

\section{The dipole cross-section}
\subsection{Balitsky-Kovchegov equation}
\label{sec:BK}
All single inclusive particle production cross sections in the Color
Glass Condensate formalism depend on the Fourier transform $C
(l_\perp)$ of the dipole cross section\footnote{Note that in case of
gluon scattering one has a dipole in the adjoint representation.},
which is the probability for scattering of a high energy
quark-antiquark dipole on a target. The dipole cross section evolves
with rapidity according to the JIMWLK equation, so that given the
dipole profile at some initial point $x_0$, the dependence of the
cross section on $x$ (or rapidity) can be in principle obtained from
the JIMWLK equation. In practice, this is very difficult to do (see
\cite{RummuW1}) and it is much simpler to use an approximate form of
this evolution -- known as the BK equation \cite{Balit1,Kovch3} --
which can be obtained from the JIMWLK equation as a mean field
approximation in the large $N_c$ limit. Here we will use the numerical
solution of the BK equation in order to investigate the properties of
the Drell-Yan structure functions.

In momentum space, the BK equation takes the form
\begin{equation}
\frac{\partial {{T}}(k_\perp,Y)}{\partial Y}
=
\overline{\alpha}_s (K\otimes {T})(k_\perp,Y)
-
\overline{\alpha}_s {T}^2(k_\perp,Y)\; ,
\label{eq:kov}
\end{equation}
where we denote $\overline{\alpha}_s\equiv \alpha_s N_c/\pi$.  The
operator $K$ is the well known BFKL kernel in momentum
space~\cite{KuraeLF1,BalitL1}, whose action on the function
${T}$ is given by
\begin{equation}
(K\otimes {T})(k_\perp,Y)
\equiv
\int\limits_0^{+\infty}
\frac{d(k^{\prime 2}_\perp)}{k^{\prime 2}_\perp}\;
\left\{
\frac{k^{\prime 2}_\perp{T}(k^\prime_\perp,Y)
-k_\perp^2{T}(k_\perp,Y)}{\big|k_\perp^2-k^{\prime 2}_\perp\big|}
+
\frac{k_\perp^2{T}(k_\perp,Y)}{\sqrt{4k_\perp^{\prime 4}+k_\perp^4}}
\right\}\; .
\end{equation}
  The function ${T}(k_\perp,Y)$ is the Bessel-Fourier
transform of the dipole-target scattering amplitude $T(r_\perp,Y)$:
\begin{equation}
{T}(k_\perp,Y)
=
\int\limits_0^{+\infty} \frac{dr_\perp}{r_\perp}  \, 
J_0(k_\perp r_\perp) \,  T(r_\perp,Y)\; ,
\label{eq:unint-gluon}
\end{equation}
where $r_\perp$ is the size of the $q\bar{q}$ dipole and $k_\perp$ is
its conjugate transverse momentum.  The dipole amplitude $T$ is
defined in terms of the correlator of two Wilson lines of gauge fields
in the target as 
 \begin{equation}
 T(r_\perp,Y)
=
1- \frac{1}{N_c}{\rm Tr} 
\left<{ V}^\dagger(0) { V}(\r_\perp)\right>_{_Y} \; ,
 \label{eq:dipole}
 \end{equation}
where we have assumed translation invariance in the transverse plane
in order to set the quark transverse coordinate to $0$. The scattering
amplitude $T$ is related to the function $C(\l_\perp)$ by
\begin{equation}
  C(\l_\perp)\equiv \int d^2\r_\perp
  \;e^{i\l_\perp\cdot \r_\perp}\; \left[1-T(\r_\perp,Y)\right]\; .
\end{equation}
Therefore, in order to obtain the function $C(\l_\perp)$ needed in the
evaluation of the Drell-Yan structure functions, we must first solve
the BK equation \ref{eq:kov} for $T(k_\perp,Y)$, then invert
eq.~(\ref{eq:unint-gluon}) in order to obtain $T(r_\perp,Y)$, and
finally obtain $C(\l_\perp)$ by a Fourier transform.

The BK equation must be supplemented by an initial condition at $Y=0$,
for which we take the value of $T(k_\perp)$ predicted by the MV model
\cite{McLerV1,McLerV2,McLerV3} with an initial saturation momentum of
$Q_s^2=2$~GeV${}^2$. This initial condition is assumed to correspond
to an $x_2$ in the nucleus of $x_2^0\equiv 10^{-2}$, which means that
the parameter $Y$ in $T(r_\perp,Y)$ should be interpreted as
$Y=\ln(x_2^0/x_2)$. For values of $x_2$ larger than $x_2^0$, we use
the following naive extrapolation~:
\begin{equation}
C(\l_\perp,x_2)=\left(\frac{1-x_2}{1-x_2^0}\right)^4\;C(l_\perp,x_2^0)\; .
\label{eq:extra}
\end{equation}
The idea behind this extrapolation is that the unintegrated gluon
distribution that appears in the gluon yield (see \cite{BlaizGV1}) is
proportional to $C(l_\perp)$, and should vanish like $(1-x_2)^4$ when
$x_2$ approaches $1$ (the exponent $4$ comes from quark counting rules).

Note also that we solve the BK equation with a fixed value of the
coupling constant $\alpha_s$. However, we set it to a rather low value
in order to correctly reproduce the rate of growth of the saturation
scale, as inferred for instance from the study of geometrical scaling
at HERA \cite{StastGK1}. This approach has already led to good results
in the analysis of limiting fragmentation in \cite{GelisSV1}.

\subsection{Models of the dipole cross section}
As an alternative to solving the BK equation, one can model the
behavior of the dipole cross section based on its known
properties. These phenomenological models have been used to
investigate particle production at RHIC and LHC as well as structure
functions in Deeply Inelastic Scattering in HERA. Here we will use two
of these models, known as the KKT \cite{KharzKT2} and DHJ
\cite{DumitHJ1,DumitHJ2} parameterizations, to calculate rapidity and
transverse momentum dependence of the Drell-Yan structure functions in
the kinematics appropriate for RHIC and LHC.

The KKT parameterization of the dipole cross section was quite
successful in description of the forward rapidity hadron production
cross section in deuteron-gold collisions at RHIC. In this
parameterization, the dipole forward amplitude is modeled as
\begin{eqnarray} 
T(\r_\perp,Y) = 1-\exp\left[ - \frac{1}{4} [r_\perp^2
  Q_s^2(Y)]^{\gamma(r_\perp,Y)}\right]\; .
  \label{eq:cs_kkt}
\end{eqnarray}
In the KKT parameterization, the anomalous dimension
$\gamma(r_\perp,Y)$ is given by
\begin{eqnarray}
\gamma(r_\perp,Y) = \frac{1}{2}\left(1+
\frac{\xi(r_\perp,Y)}{\xi(r_\perp,Y)+\sqrt{2\xi(r_\perp,Y)}+28\zeta(3)}
   \right)
   \label{eq:ano_kkt}
\end{eqnarray}
and 
\begin{equation}
\xi (r_\perp,Y) \equiv \frac{\log(1/r_\perp^2 Q_0^2)}{(\lambda/2)(Y-Y_0)}~.
\end{equation}
The saturation scale is given by $Q_s(Y) = Q_0 \exp [\lambda
(Y-Y_0)/2] $ with $Y = \ln 1/x_2$ and $Y_0 = 0.6, \, \lambda =
0.3$. While the KKT parameterization of the dipole cross section was
the first one to reproduce quantitatively the forward rapidity RHIC
data, it has some shortcomings. For example, the assumed dependence of
the anomalous dimension on $r_\perp$ (or equivalently $k_\perp$) is
too flat as shown in \cite{DumitHJ1}.

In the DHJ parameterization \cite{DumitHJ1,DumitHJ2} of the dipole
cross section, the anomalous dimension is given instead by
\begin{equation}
\gamma ( r_\perp,Y) \equiv \gamma_s + 
 (1 - \gamma_s) 
\frac{\log (1/r_\perp^2 \, Q_s^2)}{\lambda \, Y +  \log (1/r_\perp^2 \, Q_s^2) + d \, \sqrt{Y}}
\label{eq:del_gam}
\end{equation}
with $\gamma_s \simeq 0.62$, and $d=1.2$. This model has the advantage
that it accounts for the more rapid change of the anomalous dimension
with transverse size (or momentum) which is needed in order to
quantitatively describe the hadron transverse momentum spectra in both
forward and mid rapidity deuteron-gold collisions at RHIC. For further
details of the two models and their applications to RHIC data, we
refer the reader to \cite{DumitHJ1,DumitHJ2}. In the following we do
not consider the KKT model any further, and we use either the BK
evolution equation or the DHJ model.

\section{Numerical results}
\label{sec:results}
\subsection{Kinematics}
Let us complete the eqs.~(\ref{eq:T-all}) by some useful formulas for
the denominators that appear in the ${\cal T}_\alpha$'s~:
\begin{eqnarray}
&&
M^2-2p\cdot k= M^2-\frac{M^2+k_\perp^2}{z}\; ,
\nonumber\\
&&
M^2+2q\cdot k =\frac{M^2}{z}
+
\frac{(z\q_\perp-(1-z)\k_\perp)^2}{z(1-z)}
\; ,
\end{eqnarray}
where we recall that the variable $z$ is defined as $z=k^-/p^-$. These
formulas highlight the fact that the ${\cal T}_\alpha$'s defined in
eqs.~(\ref{eq:T-all}) depend on the parameters $M,\k_\perp,y$ of the
virtual photon, on the transverse momentum $\q_\perp$ of the outgoing
quark, and on the momentum fraction $z$ of the photon relative to the
incoming quark. In particular, this variable $z$ is the only reference
to the momentum of the incoming quark in these formulas.

For an incoming quark carrying the fraction $x_1$ of the longitudinal
momentum of the proton, we also have~:
\begin{equation}
zx_1=\sqrt{\frac{M^2+k_\perp^2}{s}}\;e^{-y}\; .
\end{equation}
Note that since both $z$ and $x_1$ must be smaller than unity, this
implies obvious limits to the phase-space allowed for the produced
virtual photon, since we must have~:
\begin{equation}
x_1^{\rm min}\equiv
\sqrt{\frac{M^2+k_\perp^2}{s}}\;e^{-y}\le 1\; .
\end{equation}
The notation for this quantity comes from the fact that it is also the
minimal allowed value of $x_1$. Thus, the structure function
$T_\alpha$ for the proton-nucleus system can be written as
\begin{equation}
T_\alpha^{\rm pA}
=\int\limits_{x_1^{\rm min}}^1dx_1\;q(x_1,\mu^2)\;
\frac{\pi R_{_A}^2}{2(1-\frac{x_1^{\rm min}}{x_1})}
\int\frac{d^2\q_\perp}{(2\pi)^2}\,C(\q_\perp+\k_\perp,x_2)
\;{\cal T}_\alpha\left(z=\frac{x_1^{\rm min}}{x_1}\right)\; .
\label{eq:T-x1}
\end{equation}
One last point remains to be clarified: the dependence of the function
$C$ on $x_2$ is now explicitly shown. This dependence arises from the
renormalization group resummation of small $x$ effects via JIMWLK equations
and has the interpretation as the longitudinal momentum fraction at which 
the nucleus is probed. It can be evaluated as
\begin{equation}
x_2
\equiv 
\frac{k^++q^+}{\sqrt{s/2}}
=
x_1^{\rm min}e^{2y}
+
\frac{q_\perp^2}{s(x_1-x_1^{\rm min})}\; .
\end{equation}
Note that the minimal value of $x_2$ is 
\begin{equation}
x_2^{\rm min}
=
\sqrt{\frac{M^2+k_\perp^2}{s}}\;e^{+y}\; ,
\end{equation}
(With our choice to have the nucleus moving in the $+z$ direction,
small values of $x_2$ are reached when the photon rapidity is
negative.) and that the relation between $x_1$ and $x_2$ is more
symmetric if written as~:
\begin{equation}
(x_1-x_1^{\rm min})(x_2-x_2^{\rm min})=\frac{q_\perp^2}{s}\; .
\end{equation}
Moreover, we can if necessary replace the integration over $x_1$ in
eq.~(\ref{eq:T-x1}) by an integration over $x_2$ by noting that
\begin{equation}
\frac{d\, T_\alpha^{\rm pA}}{dx_2 d^2\q_\perp}
=
\frac{x_1-x_1^{\rm min}}{x_2-x_2^{\rm min}}\;
\frac{d\, T_\alpha^{\rm pA}}{dx_1 d^2\q_\perp}\; .
\label{eq:x2-x1}
\end{equation}
This is useful in order to determine what are the values of $x_2$ that
contribute the most to the Drell-Yan structure functions.

\subsection{$x_1$ and $x_2$ dependence of the integrand}
Let us first study the $x_1$ and $x_2$ dependence of the integrand. We
do this for pA collisions at the LHC center of mass energy,
i.e. $\sqrt{s}=8.8$~TeV, for nuclei with $A^{1/3}=6$.
\begin{figure}[htb]
\begin{center}
\resizebox*{10cm}{!}{\rotatebox{-90}{\includegraphics{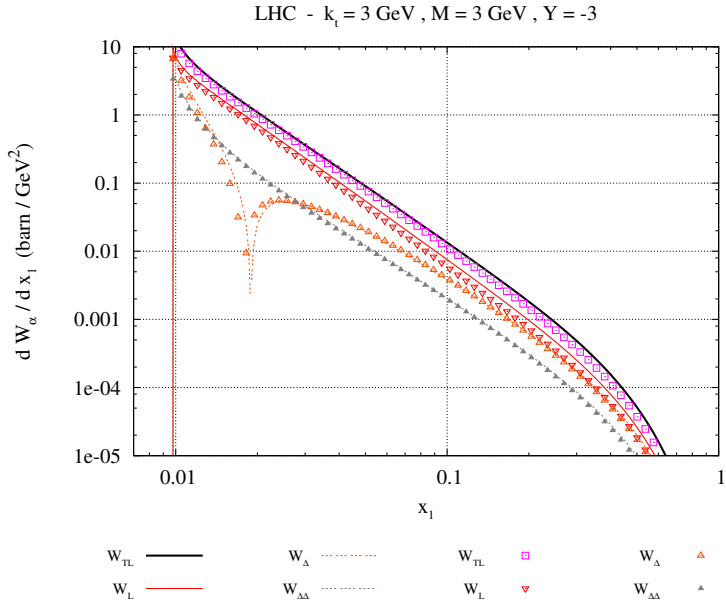}}}
\end{center}
\caption{\label{fig:lhc_x1} $x_1$ dependence of the integrand for the
$W$'s. The kinematical parameters of the virtual photons are
$k_\perp=3$~GeV, $M=3$~GeV and $y=-3$, and the center of mass energy
is $\sqrt{s}=8.8$~TeV.  Lines: BK evolution. Points: DHJ model.}
\end{figure}
In figure \ref{fig:lhc_x1}, we display the $x_1$ dependence for the
four $W$'s (i.e. we have multiplied eq.~(\ref{eq:T-x1}) by the
rotation matrix $R_{_{CS}}$ given in eq.~(\ref{eq:Rcs})), for
$k_\perp=M=3$~GeV and $y=-3$. We use the CTEQ6 (see \cite{KretzLOT1}
for details) set of parton distributions for the quark and antiquark
distributions of the proton. In this figure, we compare the $x_1$
dependence that results from the BK evolution of the dipole
cross-section, and from the DHJ model. As one can see, the two models
lead to results that are very similar: the largest contributions come
from values of $x_1$ immediately above the minimal value $x_1^{\rm
min}$, and the integrand drops very quickly as $x_1$ increases. One
can also note that the integrand for $W_{_\Delta}$ vanishes and
changes sign at some intermediate value of $x_1$ (for this integrand,
we have displayed its absolute value, hence the cusp at the location
of the zero on the logarithmic plot). In fact, for the kinematical
parameters of the virtual photon used here, the two descriptions of
the dipole cross-sections are very close even quantitatively. However,
as we shall see later, this is not the case everywhere.

\begin{figure}[htb]
\begin{center}
\resizebox*{10cm}{!}{\rotatebox{-90}{\includegraphics{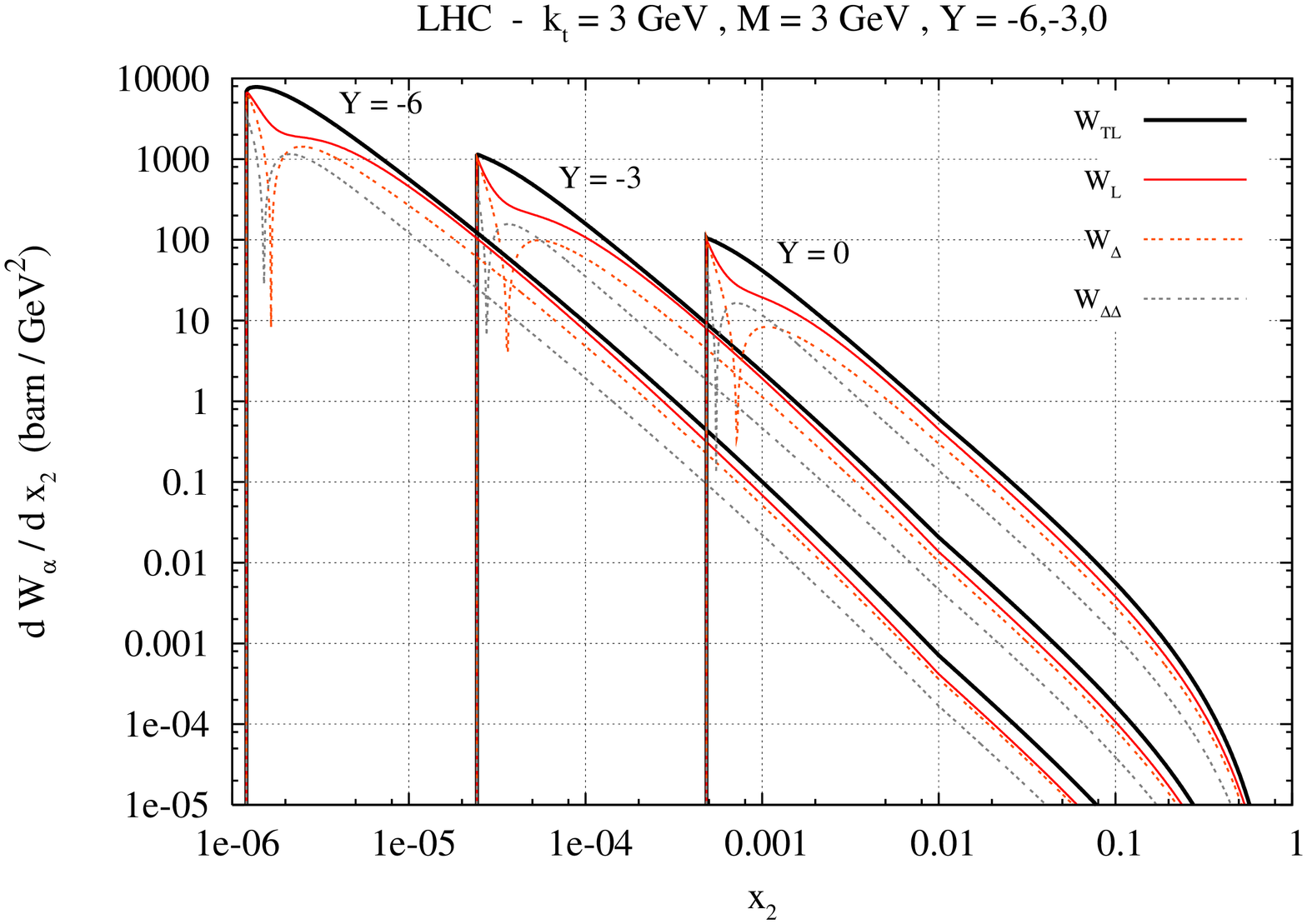}}}
\end{center}
\caption{\label{fig:lhc_x2} $x_2$ dependence of the integrand for the
$W$'s, with the BK evolution. The kinematical parameters of the
virtual photons are $k_\perp=3$~GeV, $M=3$~GeV and we consider the
rapidities $y=-6,-3$ and 0. The center of mass energy is
$\sqrt{s}=8.8$~TeV.}
\end{figure}

In figure \ref{fig:lhc_x2}, we have used the eq.~(\ref{eq:x2-x1}) in
order to display the $x_2$ dependence of the same integrands. In order
to see how the dominant values of $x_2$ vary with the photon rapidity,
we have considered three values of $y$, $y=-6, -3$ and $0$ (the first
one can be reached only by the CMS and ATLAS detectors, while the last
two may also be studied by ALICE). To avoid overcrowding the figure,
we display the results of this study only in the case of the BK
evolution\footnote{At $x_2=10^{-2}$, one can notice a very small
discontinuity in the slope of all the curves. This is the value of
$x_2$ where we switch to the extrapolation described by
eq.~(\ref{eq:extra}). As long as this point is far from $x_2^{\rm
min}$, this extrapolation will not affect significantly the result of
the integration over $x_2$.}.
As in the case of the $x_1$ dependence, the dominant values of $x_2$
are located just above the minimal value $x_2^{\rm min}$, and the
integrand decreases quickly as $x_2$ increases. This enables one to
select different windows in $x_2$ by varying the kinematical
parameters of the virtual photon. This is illustrated in figure
\ref{fig:lhc_x2} with three different values of the photon
rapidity. As one can see, it is possible to explore very small values
of $x_2$ by going to negative rapidities, provided that the photon
transverse momentum and invariant mass remain rather small.

\subsection{Structure functions for pA and dA collisions}
\subsubsection{$k_\perp$ dependence}
Let us now display the results for the $W$'s, integrated over the
momentum fraction $x_1$ of the incoming quark. We have performed this
calculation both for the LHC and for RHIC. For the latter, we have
used a center of mass energy $\sqrt{s}=200$~GeV and the incoming
projectile is taken to be a deuteron\footnote{The CTEQ package does
not provide parton distributions for the deuteron. In order to
overcome this restriction, we assume that one can just add the parton
distributions of a proton and of a neutron, i.e. we neglect the
shadowing that may occur in the deuteron. This is a rather good
assumption as long as $x_1$ is not too small. In order to obtain the
quark distributions in the neutron, we just exchange the $u$ and $d$
valence distributions of a proton, and we assume that the sea quark
distributions are identical in a proton and a neutron.} rather than a
proton.
\begin{figure}[htb]
\begin{center}
\resizebox*{6cm}{!}{\rotatebox{-90}{\includegraphics{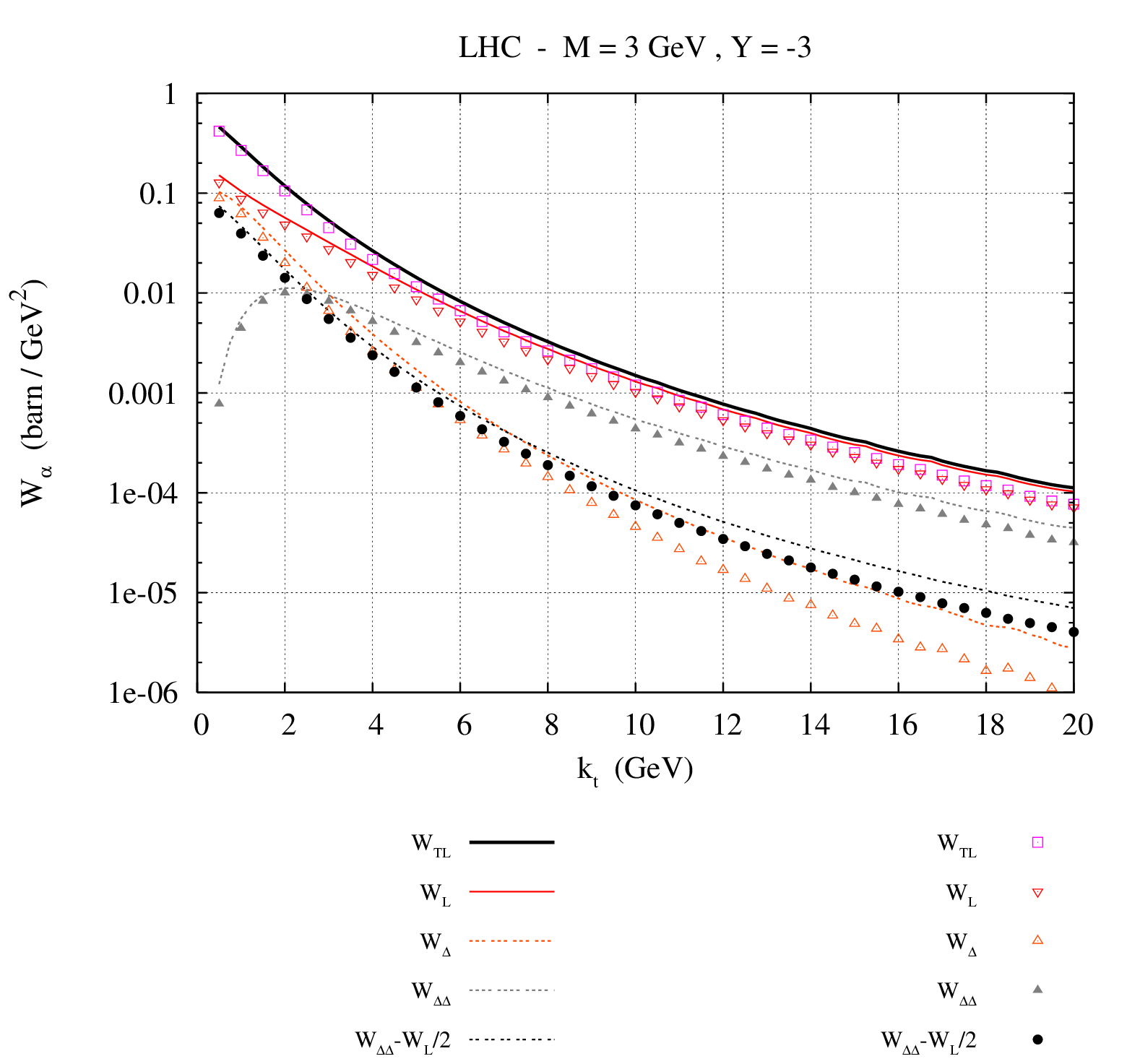}}}
\resizebox*{6cm}{!}{\rotatebox{-90}{\includegraphics{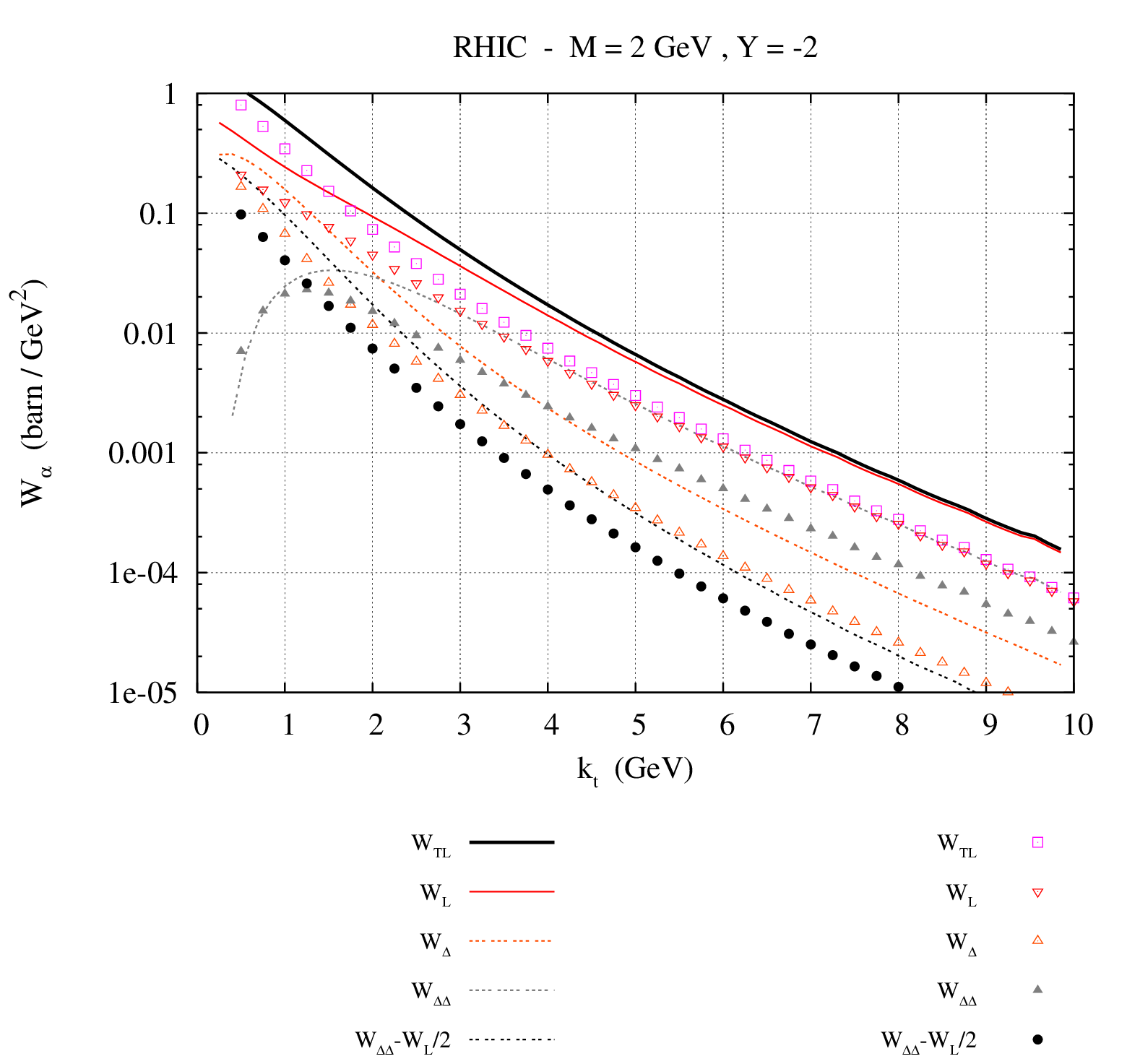}}}
\end{center}
\caption{\label{fig:K}Drell-Yan structure functions as a function of
$k_\perp$, for RHIC (right) and LHC (left). Lines: BK
evolution. Points: DHJ model.}
\end{figure}

In figure \ref{fig:K}, we display the dependence of the Drell-Yan
structure functions on the transverse momentum of the virtual photon,
for a fixed invariant mass and rapidity (this rapidity is chosen
negative to be in the region where the nucleus is explored at small
$x_2$). The calculation has been performed with the function
$C(\l_\perp,x_2)$ resulting from the BK evolution (lines) and from the
DHJ model (points), both for the LHC (left) and for RHIC (right). We
also display on the same plot the difference
$W_{_{\Delta\Delta}}-W_{_L}/2$, which is expected to vanish if the
Lam-Tung relation is satisfied. Since we have checked explicitly that
this combination vanishes identically in the leading twist
approximation, any non-zero value for this quantity is due to higher
twist corrections -- i.e. to multiple scatterings of the quark in the
nuclear target.
\begin{figure}[htb]
\begin{center}
\resizebox*{6cm}{!}{\rotatebox{-90}{\includegraphics{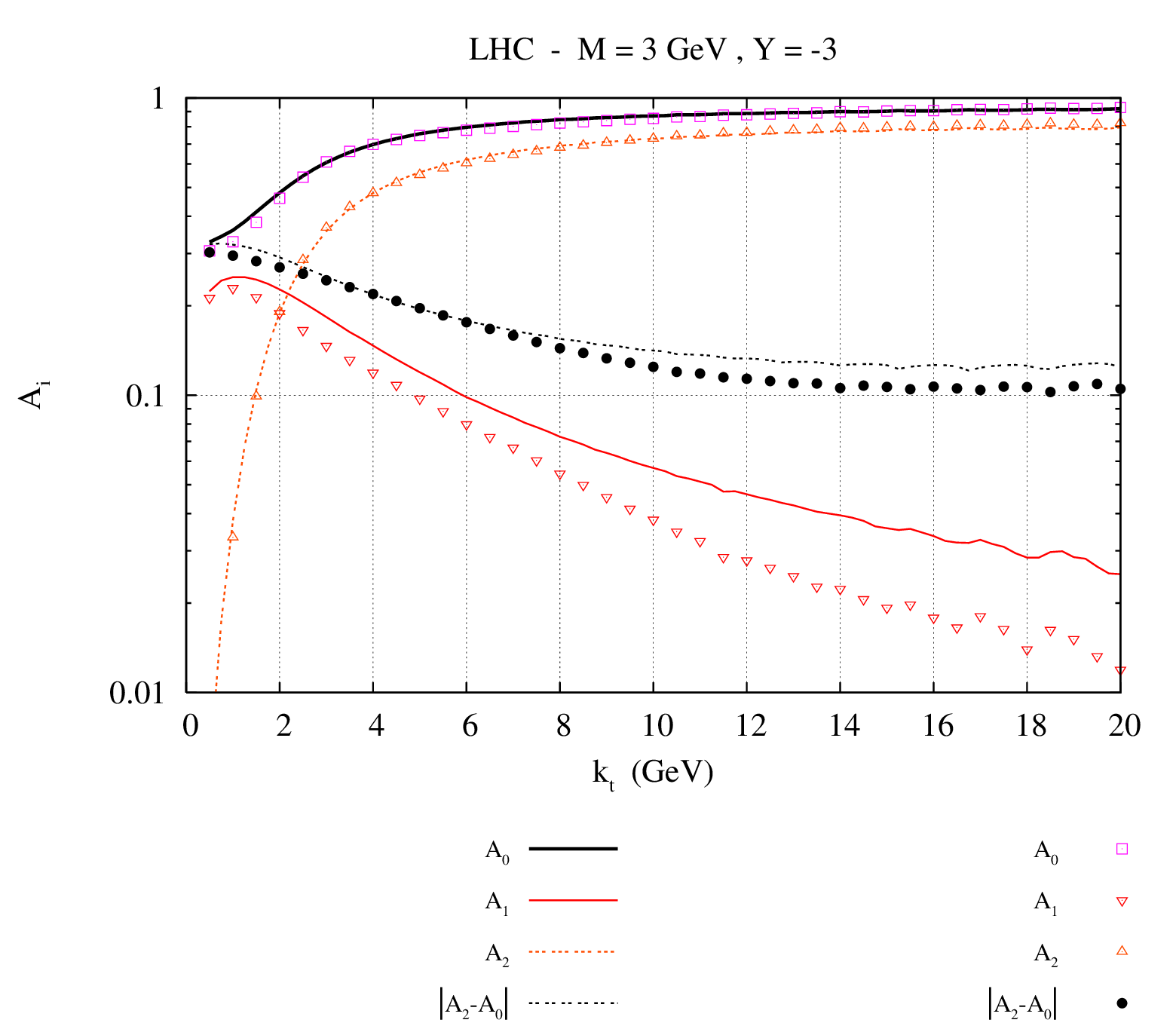}}}
\resizebox*{6cm}{!}{\rotatebox{-90}{\includegraphics{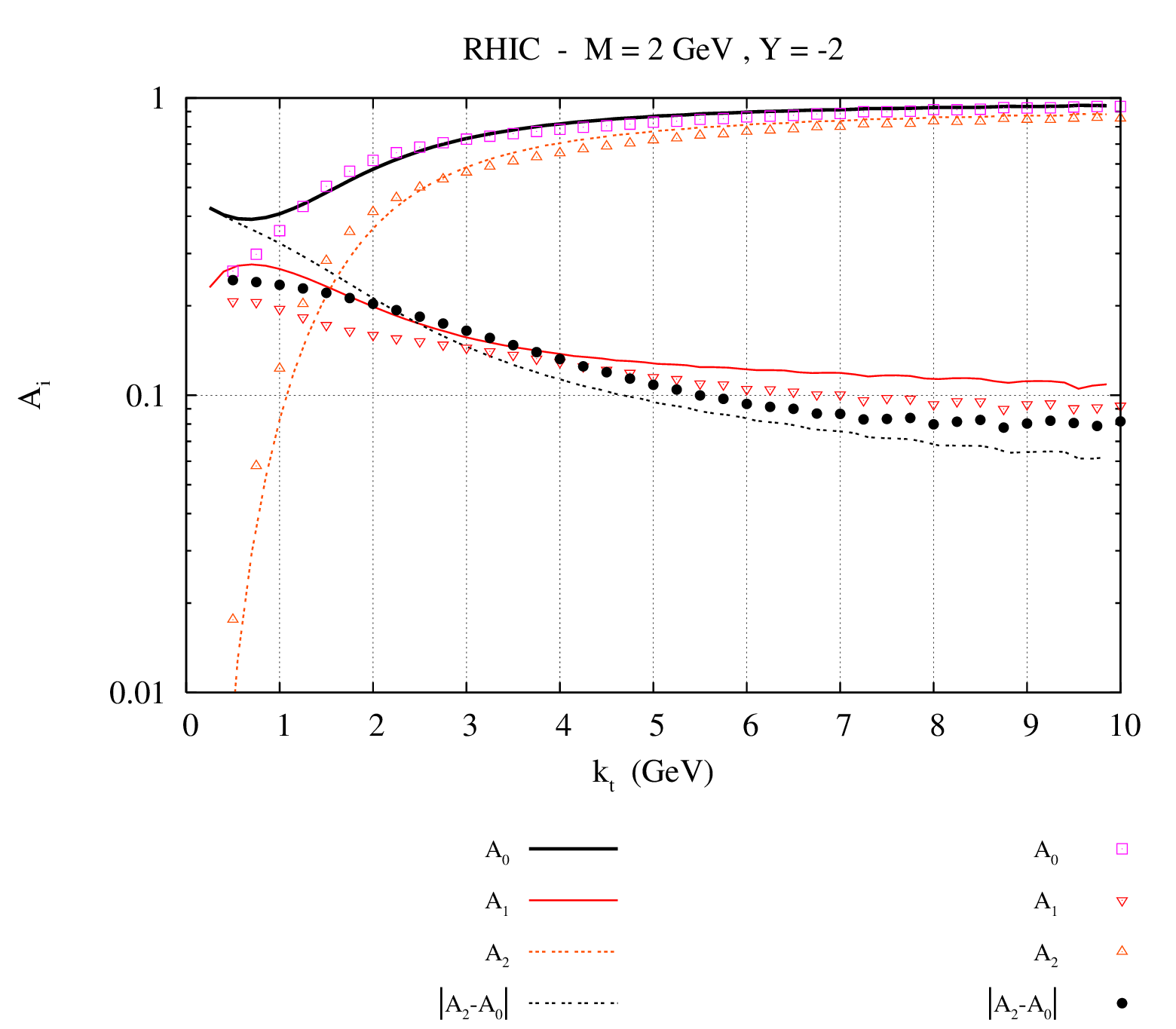}}}
\end{center}
\caption{\label{fig:K_A}Angular coefficients as a function of
$k_\perp$, for RHIC (right) and LHC (left). Lines: BK
evolution. Points: DHJ model.}
\end{figure}
One can see in figure \ref{fig:K} that the BK evolution and the DHJ
model lead to qualitatively similar features in the Drell-Yan
structure functions. The difference between the two descriptions is
never very large (at most a factor 3), but one can see that it is more
pronounced at large $k_\perp$, and also more pronounced at RHIC energy
compared to LHC energy. Moreover, the fact that the $W$'s obtained
with the DHJ model drop slightly faster at large $k_\perp$ than those
obtained from the BK evolution suggests that the difference comes from
the anomalous dimension $\gamma$. In the BK evolution, it is always
slightly above $1/2$, while the DHJ parameterization is such that this
anomalous dimension goes back to unity at large momentum.

Note that the difference $W_{_{\Delta\Delta}}-W_{_L}/2$ should be
compared to the structure function $W_{_{TL}}$ (indeed, the
$W_{_{TL}}$ structure function, obtained by contracting with
$g^{\mu\nu}$, is closely related to the {\sl total}
cross-section). One sees that this combination is more and more
suppressed as the transverse momentum of the virtual photon increases,
as expected for a higher twist quantity. This means that violations of
the Lam-Tung relation should be searched at low $k_\perp$ rather than
at large $k_\perp$. In figure \ref{fig:K_A}, we display the angular
coefficients defined in section \ref{sec:angular} for the same set of
parameters. Violations of the Lam-Tung relation should are visible in
the fact that the difference $A_0-A_2$ is non-zero. One can see that
they decrease, albeit very slowly, when the transverse momentum
increases.

\subsubsection{$M$ dependence}
\begin{figure}[htb]
\begin{center}
\resizebox*{6cm}{!}{\rotatebox{-90}{\includegraphics{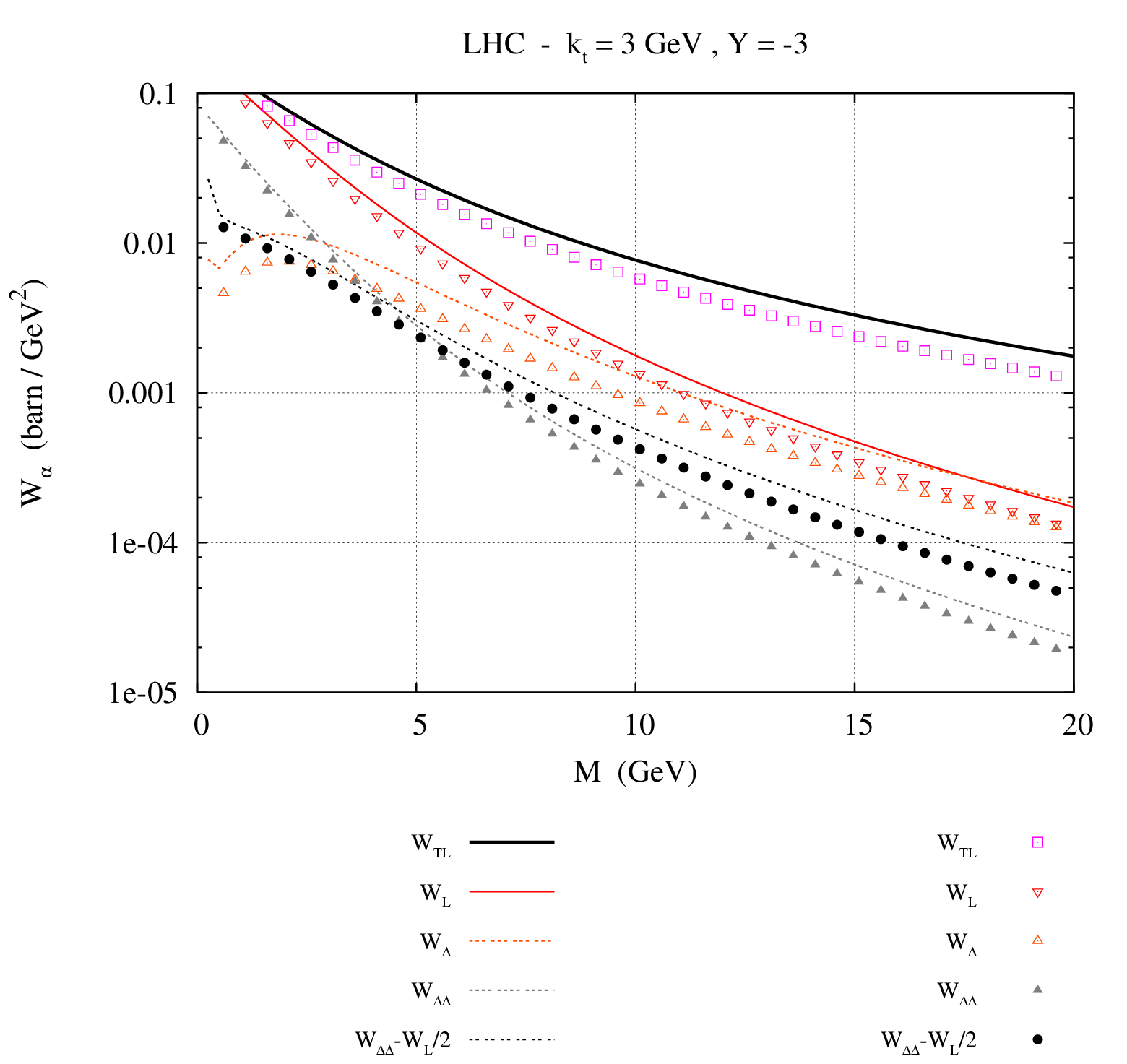}}}
\resizebox*{6cm}{!}{\rotatebox{-90}{\includegraphics{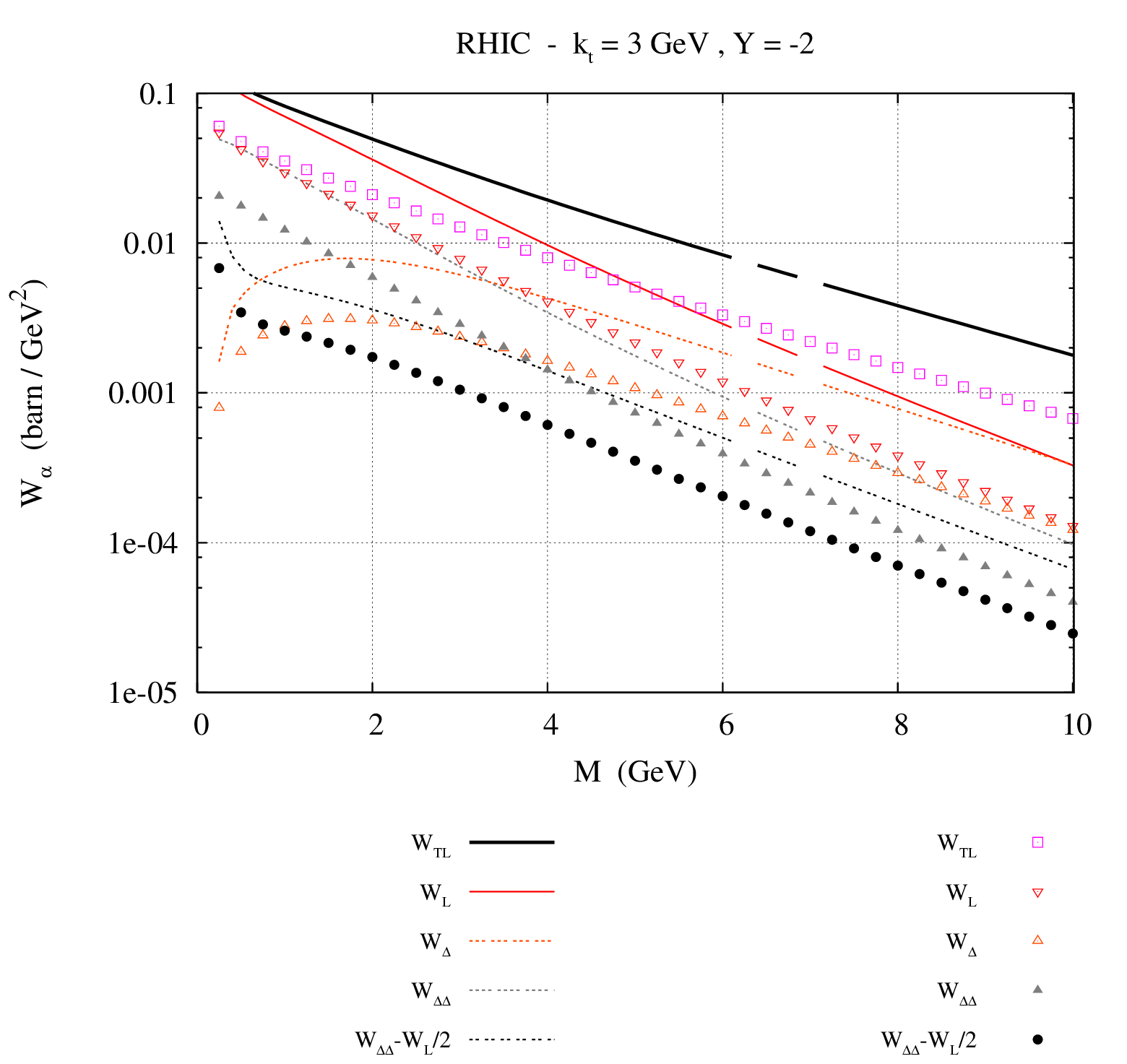}}}
\end{center}
\caption{\label{fig:M}Drell-Yan structure functions as a function of
$M$, for RHIC (right) and LHC (left). Lines: BK
evolution. Points: DHJ model.}
\end{figure}
\begin{figure}[htb]
\begin{center}
\resizebox*{6cm}{!}{\rotatebox{-90}{\includegraphics{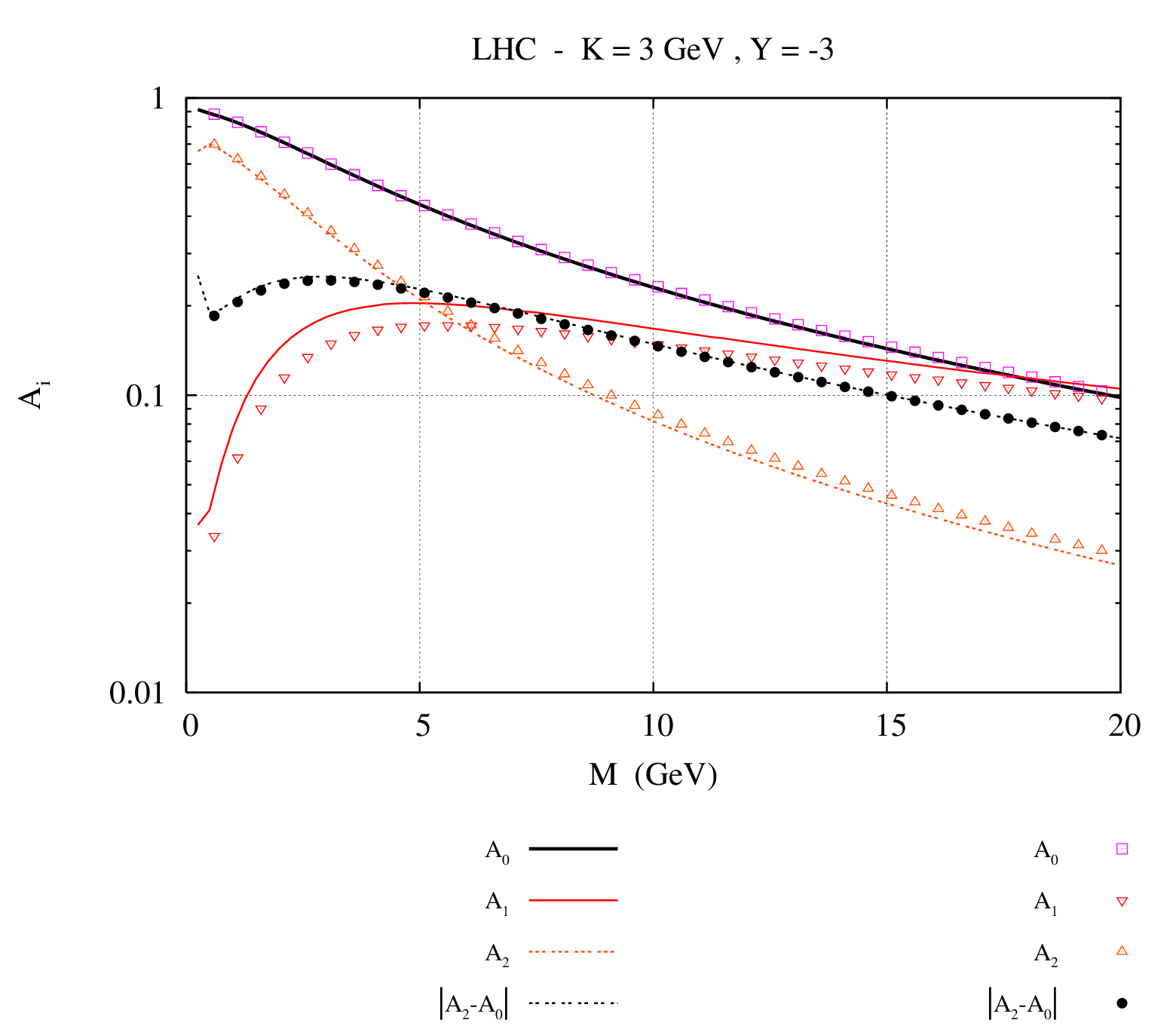}}}
\resizebox*{6cm}{!}{\rotatebox{-90}{\includegraphics{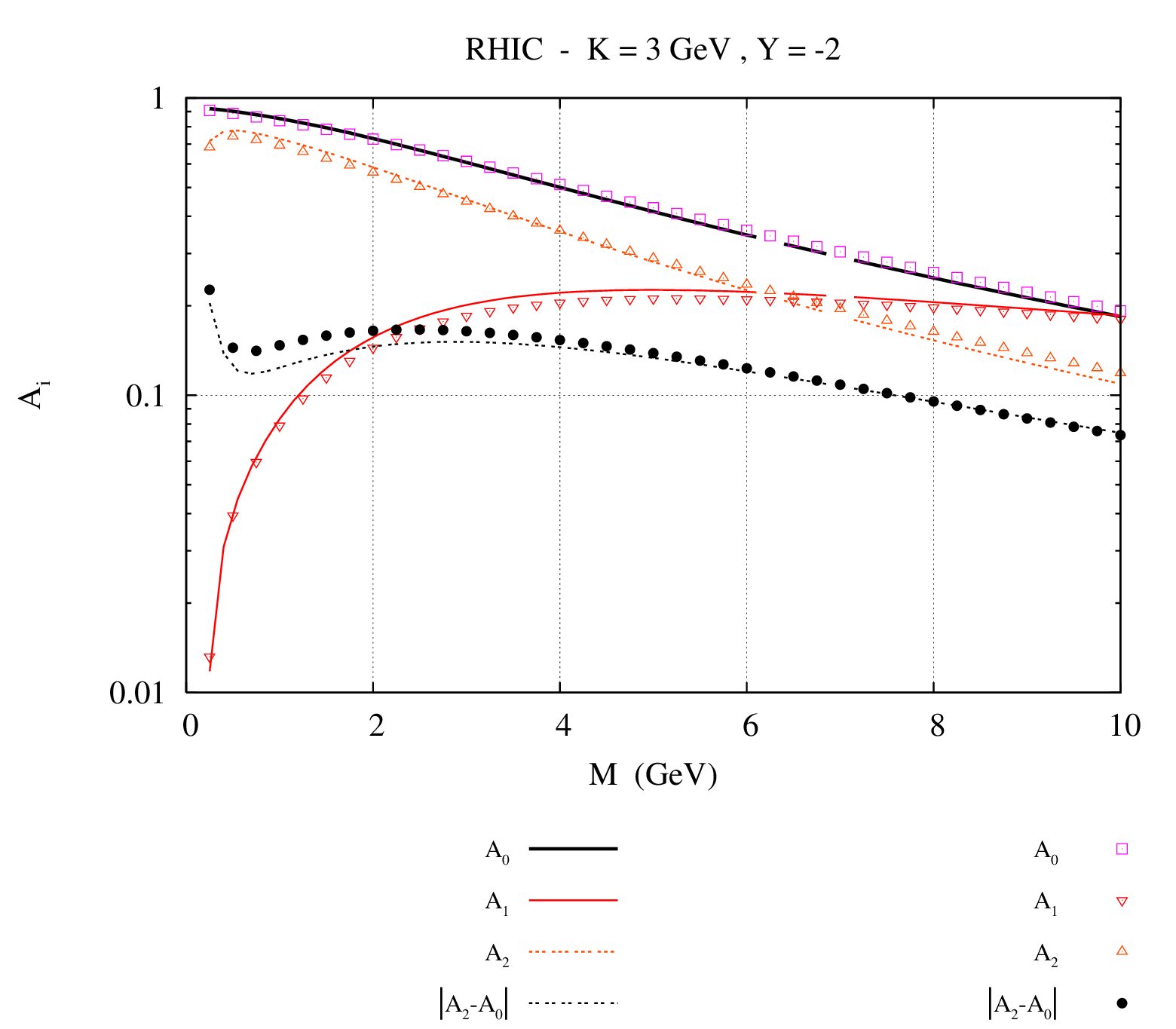}}}
\end{center}
\caption{\label{fig:M_A}Angular coefficients as a function of
$M$, for RHIC (right) and LHC (left). Lines: BK
evolution. Points: DHJ model.}
\end{figure}
We display in figures \ref{fig:M} and \ref{fig:M_A} the same structure
functions and angular coefficients as a function of the virtual photon
invariant mass. Again, the calculation was done both for RHIC and the
LHC, with the BK evolution and the DHJ model. We observe for the $M$
dependence results that are qualitatively similar to what we already
found for the $k_\perp$ dependence. The DHJ model and the BK evolution
are closer at the LHC rather than at RHIC. And again, we see that the
suppression of the difference $W_{_{\Delta\Delta}}-W_{_L}/2$ compared
to the ``total'' structure function $W_{_{TL}}$ is larger at large
mass, or conversely that violations of the Lam-Tung relation are
larger at small invariant mass.

\subsubsection{$Y$ dependence}
\begin{figure}[htb]
\begin{center}
\resizebox*{6cm}{!}{\rotatebox{-90}{\includegraphics{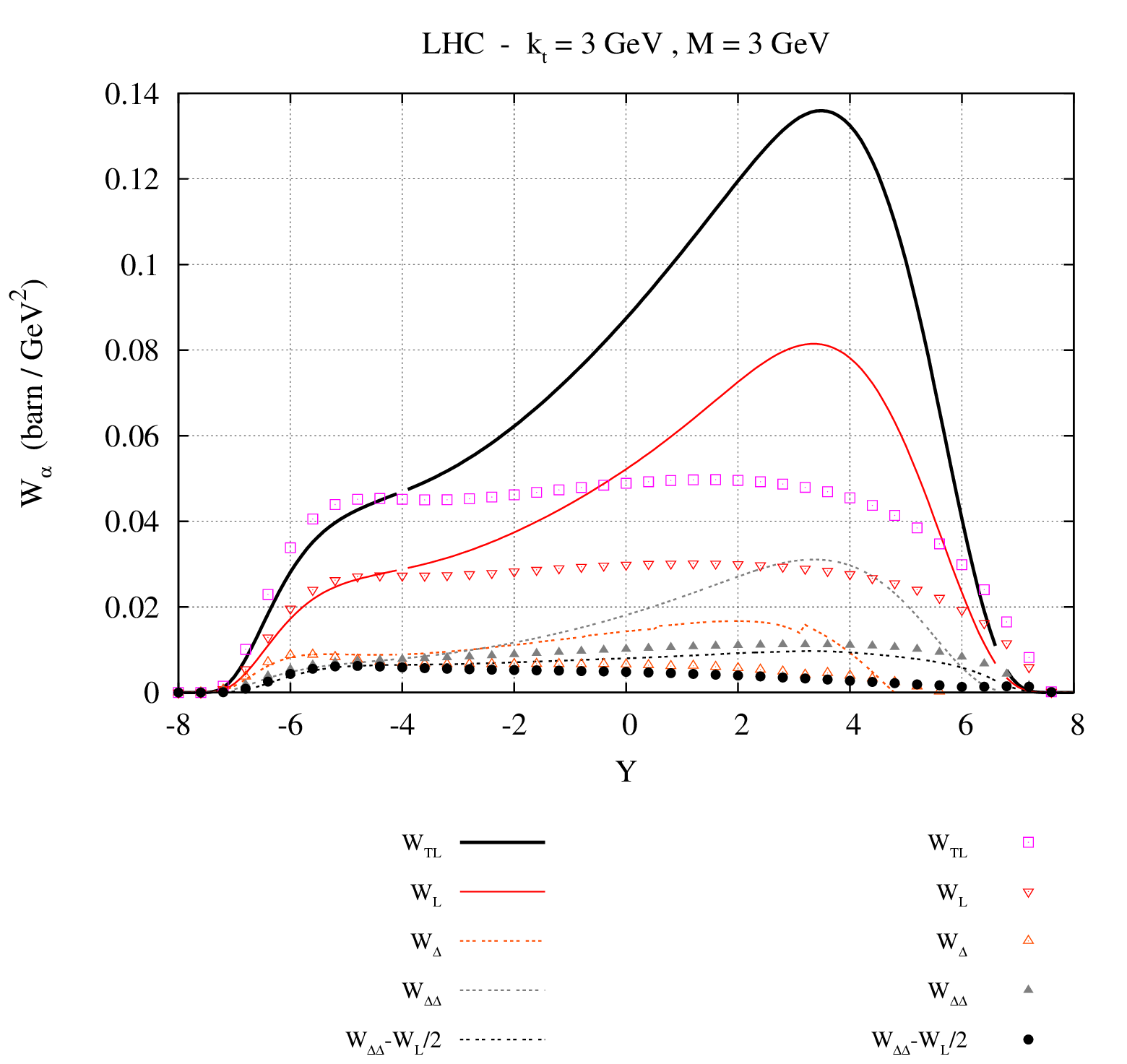}}}
\resizebox*{6cm}{!}{\rotatebox{-90}{\includegraphics{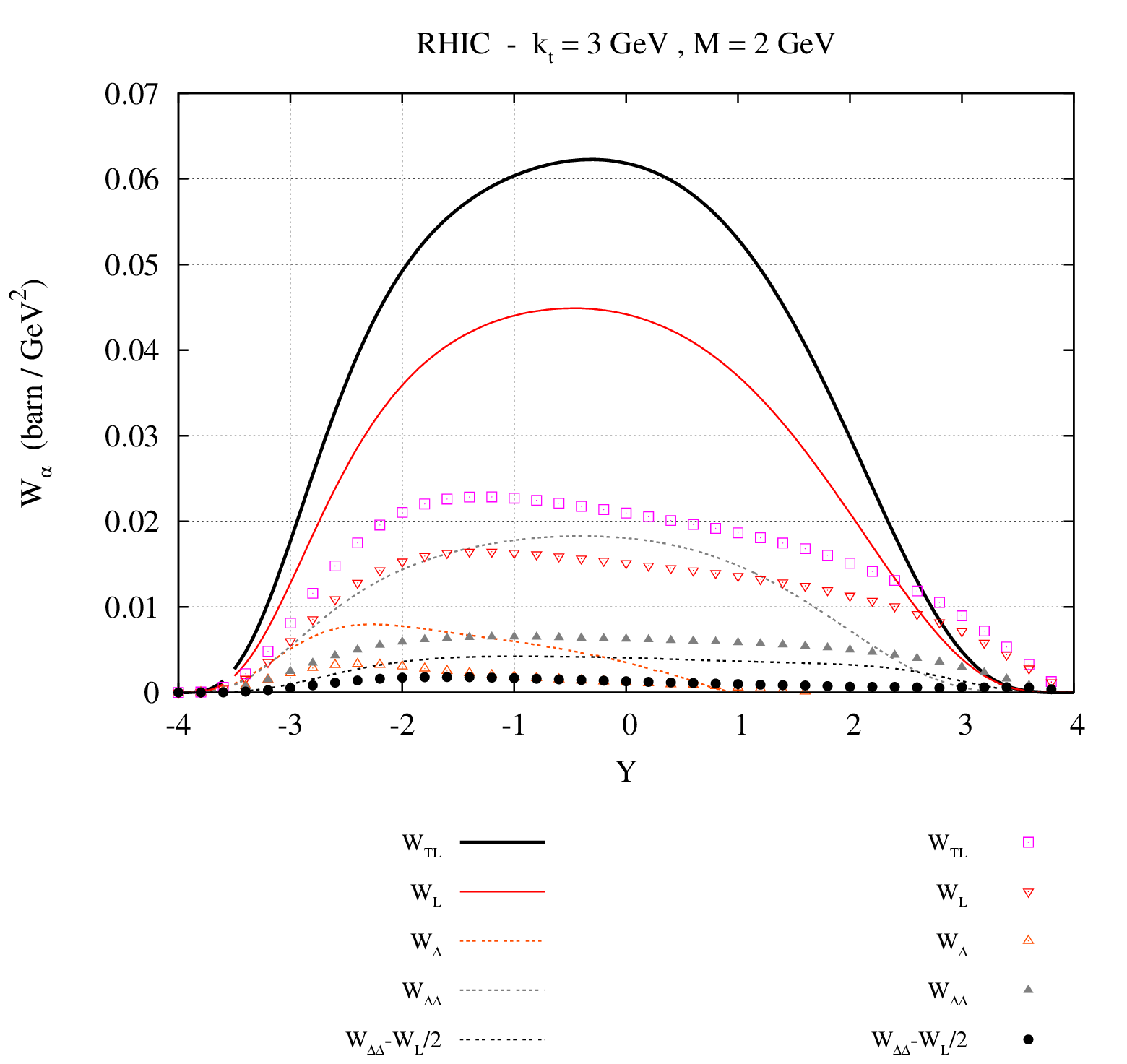}}}
\end{center}
\caption{\label{fig:Y}Drell-Yan structure functions as a function of
$Y$, for RHIC (right) and LHC (left). Lines: BK
evolution. Points: DHJ model.}
\end{figure}
\begin{figure}[htb]
\begin{center}
\resizebox*{6cm}{!}{\rotatebox{-90}{\includegraphics{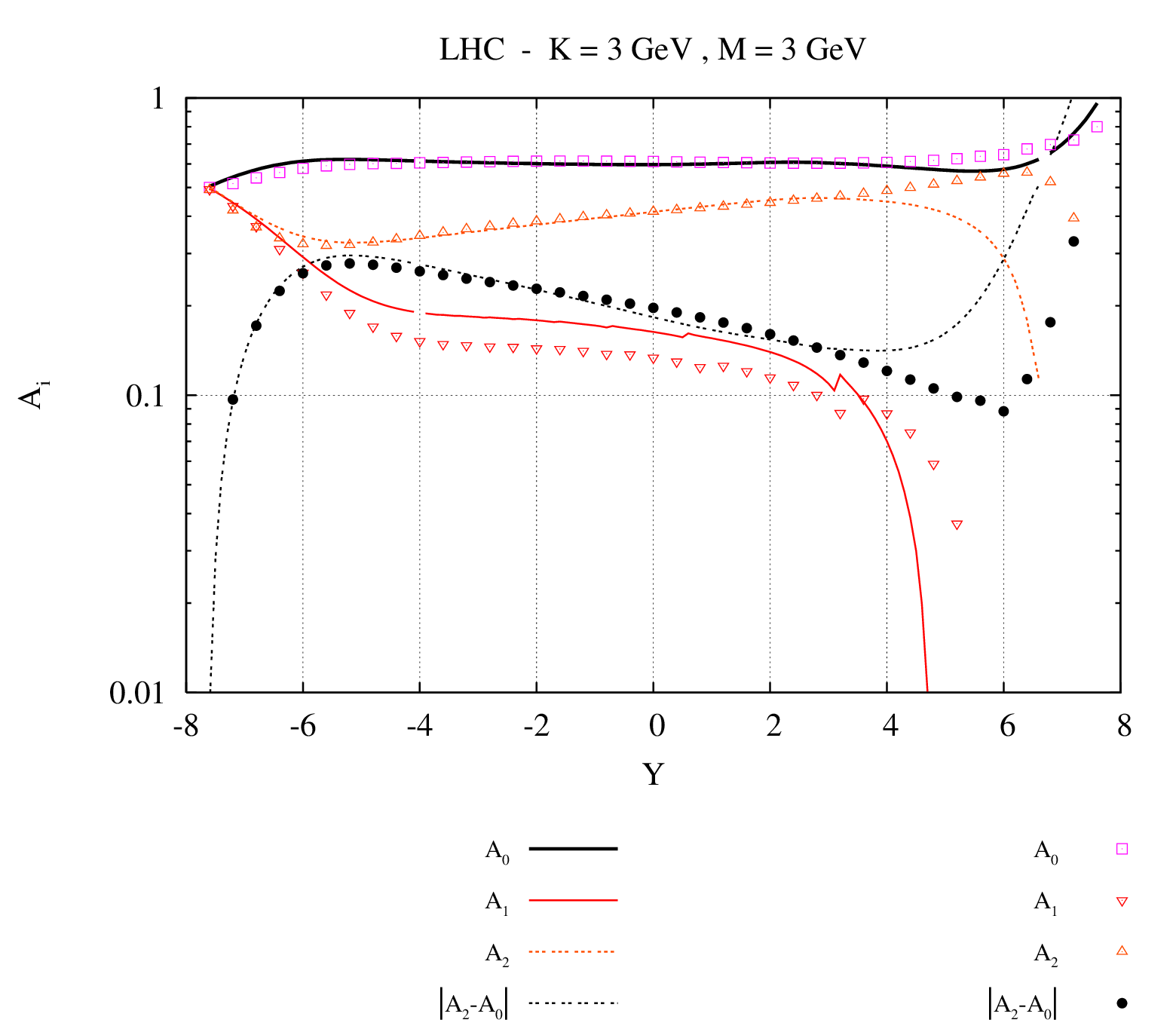}}}
\resizebox*{6cm}{!}{\rotatebox{-90}{\includegraphics{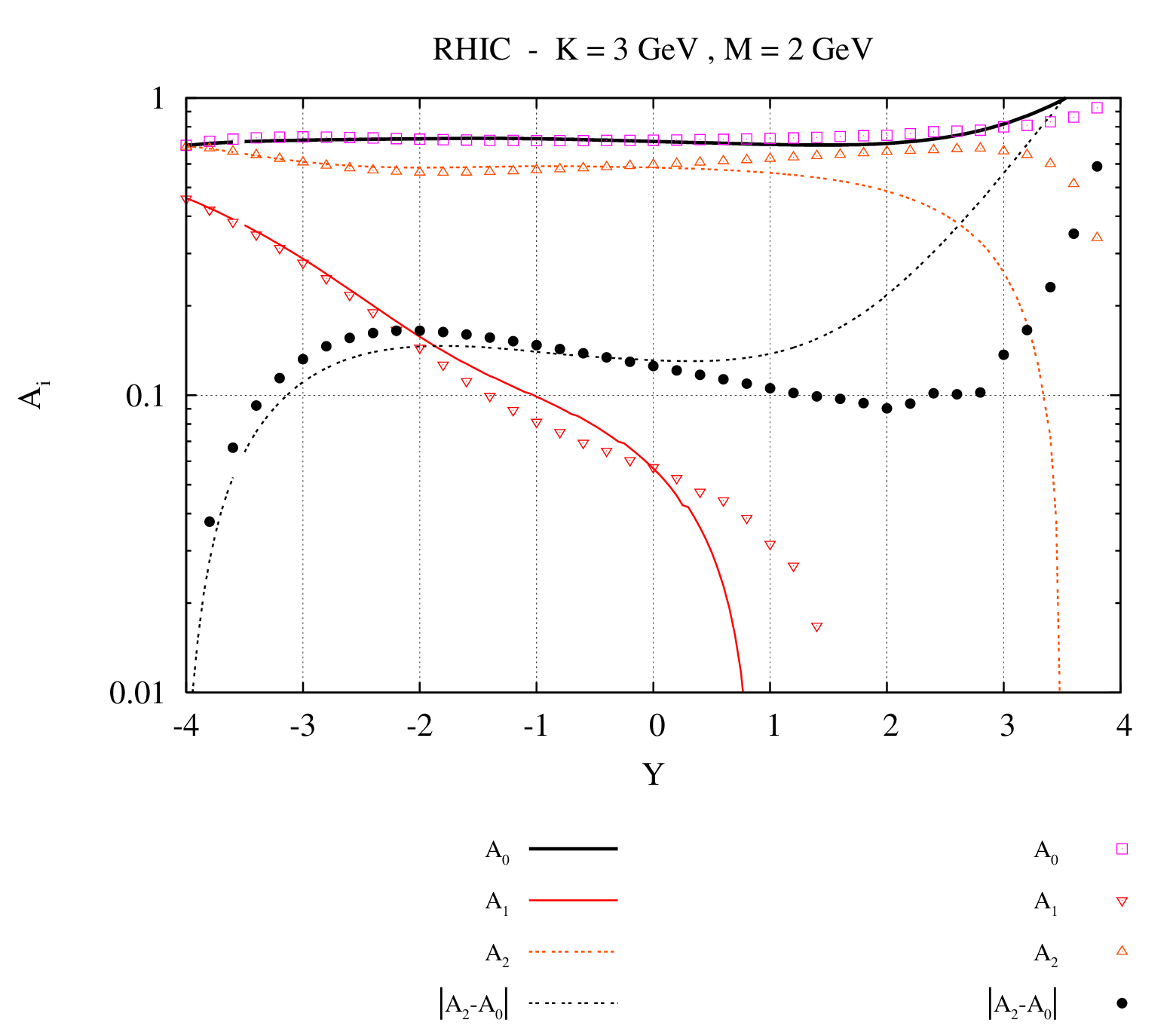}}}
\end{center}
\caption{\label{fig:Y_A}Angular coefficients as a function of
$Y$, for RHIC (right) and LHC (left). Lines: BK
evolution. Points: DHJ model.}
\end{figure}

Finally, we present results concerning the rapidity dependence, at
fixed $k_\perp$ and $M$, of the Drell-Yan structure functions and
angular coefficients in figures \ref{fig:Y} and \ref{fig:Y_A}. Note
that the vertical scale in figure \ref{fig:Y} is now linear, in order
to better show the variations with $Y$ and the differences between the
BK evolution and the DHJ model. This natural makes the differences
between the DHJ model and the BK evolution look larger than in the
previous plots, while they are in fact comparable in magnitude.

Now, one sees qualitative differences between the two descriptions of
the nuclear target. In absolute terms, the two descriptions do not
differ by vast amounts, but the variations with $Y$ are quite
different. The general trend is that the BK evolution (with the
parameters described in section \ref{sec:BK}) leads to significantly
stronger variations with the rapidity. One should keep in mind that
the DHJ model has been primarily devised to reproduce correctly the
$k_\perp$ dependence of the hadronic spectra observed by the BRAHMS
and STAR collaborations at RHIC, and in particular their slope and how
this slope varies with rapidity. However, it did not predict the
absolute normalization of these spectra, and a ``K-factor'' had to be
adjusted by hand. Moreover, the K-factor necessary to reproduce the
normalization of the data turned out to be different at each
rapidity. What we observe here may be related to this fact, and seems
to indicate that the DHJ model should still be improved in order to
get correctly the changes in the absolute normalization with rapidity.

Regardless of the description used for the function $C(\l_\perp,x_2)$
-- BK evolution or DHJ model --, one also sees very different shapes at
RHIC and the LHC. In the DHJ model, the slope tends to be inverted
from RHIC to the LHC. And with the BK evolution, one goes from fairly
symmetric shapes at RHIC to a very asymmetric rapidity distribution at
the LHC. However, one should take this remark with caution, because
our calculation of the Drell-Yan structure function treats the proton
and the nucleus in very different ways, and is only valid when $x_2$
is small in the nucleus and $x_1$ is rather large in the
proton. Therefore, the accuracy of our results at {\bf positive rapidity}
may suffer significantly from this treatment.

Finally, one can note that the violations of the Lam-Tung relation,
that one can see in figure \ref{fig:Y_A} on the difference $A_0-A_2$,
rise steadily when the photon rapidity goes from positive to negative
values (at least if one excludes the fragmentation regions of the two
projectiles). This is also expected of a higher twist effect: indeed,
at smaller $y$, i.e. at smaller $x_2$, the density of color charges in
the nucleus increases, which leads to more important rescattering
effects.

\section{Conclusions}
In this paper, we have studied in the Color Glass Condensate framework
the production of Drell-Yan pairs in proton-nucleus and
deuteron-nucleus collisions at RHIC and at the LHC. Our treatment is
asymmetric, because we only use the CGC description for the nucleus,
while the proton or deuteron is described by means of the usual parton
distributions.

All the nuclear effects that affect the Drell-Yan structure functions
are encoded in the function $C(\l_\perp)$, which is the Fourier
transform of the dipole-nucleus cross-section, and we have considered
two descriptions of this object: the DHJ model and the solution of the
BK equation (with the McLerran-Venugopalan model as the initial
condition). Overall, the most striking differences between these two
description are seen in the rapidity dependence of the structure
functions, where the BK description leads to a much stronger rapidity
dependence.

We have also discussed the Lam-Tung relation, a linear
combination of the structure functions which is known to vanish
identically up to (and including) the Next-to-Leading order in the
leading twist approximation. Such a quantity is indeed very
interesting to investigate in order to study the effects of higher twist
contributions, because any non-zero value is by definition coming from
higher twist. We observed for this quantity a behavior which is indeed
what one expects from a higher twist quantity: it drops at high
momentum or high mass slightly faster than the total cross-section,
and it is larger in regions of rapidity where the nucleus is probed at
smaller $x$, i.e. where it has a larger parton density. A detailed experimental 
study of Lam-Tung relation then should prove very fruitful in advancing our
knowledge of QCD in its novel high parton density regime.

\section*{Acknowledgements}
F.G. would like to thank the hospitality of the Institute of Nuclear
Theory where a significant part of this work has been performed, and
the financial support of the grant ACTION CNRS-NSF \#17251.  J. J-M. is
supported in part by the U.S. Department of Energy under Grant No.
DE-FG02-00ER41132.

\bibliographystyle{unsrt}
%\bibliography{biblio}

\end{document}